\documentclass[%
 aip,
 jcp,
 amsmath,amssymb,
 reprint,%
]{revtex4-1}

\usepackage{graphicx}% Include figure files
\usepackage{dcolumn}% Align table columns on decimal point
\usepackage{bm}% bold math
\usepackage[utf8]{inputenc}
\usepackage[T1]{fontenc}
\usepackage{mathptmx}
\usepackage{etoolbox}
\usepackage[ruled,lined]{algorithm2e}
\usepackage{xcolor}
\usepackage{listings}
\usepackage{newfloat}
\usepackage{graphicx}
\usepackage{threeparttable}
\usepackage{multirow}

\DeclareFloatingEnvironment[
    fileext=loc,
    name=Listing
]{code}
\makeatletter
\def\@email#1#2{%
 \endgroup
 \patchcmd{\titleblock@produce}
  {\frontmatter@RRAPformat}
  {\frontmatter@RRAPformat{\produce@RRAP{*#1\href{mailto:#2}{#2}}}\frontmatter@RRAPformat}
  {}{}
}%
\makeatother
\begin{document}

\preprint{AIP/123-QED}

\title{libNLPBE: An Open-Source Python Package for Solving the Non-Linear Poisson-Boltzmann Equation}
% Force line breaks with \\
\author{Jun-Hyeong Kim}
 \affiliation{Department of Chemistry, Duke University, Durham, North Carolina 27708, United States}
\author{Weitao Yang}%
 \email{weitao.yang@duke.edu}
\affiliation{Department of Chemistry, Duke University, Durham, North Carolina 27708, United States}
\date{\today}% It is always \today, today,

\begin{abstract}
Solvent environments surrounding a solute can significantly alter its chemical properties. These changes become more prominent in electrolyte solutions, where mobile ions largely influence the solute through electrostatic interactions. A theoretical description of these ionic effects will benefit the design of chemistry performed in electrolyte solutions. As a result, the non-linear Poisson-Boltzmann equation (NLPBE) has emerged as an efficient implicit solvent model for electronic structure calculations  to address such effects. However, the limited availability of NLPBE solvers for molecular electronic structure calculations greatly hinders theoretical investigation into the ionic effects. To improve accessibility of the NLPBE, we present libNLPBE, an open-source Python library, that solves the NLPBE for molecular systems in combination with density functional calculations, specifically to obtain electrostatic correction terms arising from the electrolyte solution environment for the Fock matrix. The library employs the density fitting (DF) approximation to efficiently calculate solute electrostatic potentials. Furthermore, we develop the modified Damped Inexact Newton Multigrid developed by Holst (mDINMH) method for solving the NLPBE. The mDINMH features a symmetric preconditioner for solving the Newton equation, which enables the use of robust multigrid methods designed for symmetric linear operators. In addition, an algebraic multigrid method has been incorporated into the mDINMH to support a broad range of grid points. We also present a GPU-accelerated version of libNLPBE to leverage parallelization efficiency of GPUs. For the evaluation of solute electrostatic potentials, the DF approach achieves $\sim$82-fold and $\sim$302-fold speedups on CPU and GPU, respectively. Furthermore, compared to the self-consistent method for solving the NLPBE, the mDINMH achieves $\sim$10-fold and $\sim$74-fold speedups on CPU and GPU, respectively. In comparison with standard solvent models, the DF+mDINMH method produces reasonable solvation free energies at zero electrolyte concentration. Furthermore, we demonstrate that, compared with the reference NLPBE solver available in Q-Chem software, the DF+mDINMH combination accurately calculates the electrostatic contribution of solvation energies, with an error up to 0.04 kcal/mol. Lastly, we show that the DF+mDINMH method is capable of accurately predicting the effects of mobile ions on the free energy of solvation. Given the efficiency of libNLPBE and its easy integration with various quantum chemistry packages, we expect the applicability of libNLPBE to a broad range of quantum chemistry platforms.
\end{abstract}

\maketitle

\section{Introduction}
Interactions between molecules and their environment play a pivotal role in controlling a wide range of chemical phenomena. For example, it is well known that solvent effects critically influence stability, reactivity, and spectroscopic properties of the solute molecule.\cite{tomasi_quantum_2005,cossi_energies_2003}
These effects become more prominent when the solvent contains electrolytes, where mobile ions in the solution give rise to interesting interfacial phenomena.\cite{klapper_focusing_1986,sarkar_electrodes_2019,westendorff_electrically_2024} Thus, elucidating the effects of ions on the solute is crucial for understanding the the shift of its chemical properties.\\
Theoretical insights into these changes would help design chemistry performed in electrolyte solutions. For example, molecular dynamics simulations could be carried out to understand how electrolytes influence the solute molecule. This approach, however, requires a simulation box densely filled with solvent molecules and electrolytes. Consequently, their degrees of freedom generate a vast number of configurations, resulting in significant computational complexity in obtaining a converged free energy of the system.\\
In contrast to explicit modeling of electrolyte solutions, implicit solvent models replace the solvent environment with a polarizable continuum, thereby eliminating the complexity of sampling the enormous configurational space generated by explicit models.\cite{pye_implementation_1999,tomasi_quantum_2005,marenich_universal_2009} Among these, the Poisson-Boltzmann equation (PBE) has emerged as an efficient implicit model for describing electrolyte solution environments.\cite{rashin_reevaluation_1985,sharp_calculating_1990,nicholls_rapid_1991,tannor_accurate_1994,letchworth-weaver_joint_2012,fisicaro_generalized_2016,ringe_function-space-based_2016,stein_poissonboltzmann_2019,herbert_dielectric_2021} The PBE treats ions in the solution as a charge density distributed over space according to Boltzmann statistics, where the associated Boltzmann factor is determined by electrostatic potentials.\\
Depending on whether the Boltzmann factor is linearized, the PBE is formulated as either the linearized (LPBE) or the non-linear Poisson-Boltzmann equation (NLPBE). The LPBE produces reasonable results under weak electrostatic potentials; however, it inherently fails to accurately describe ion distributions under strong electrostatic potentials. This limitation is evident particularly when describing solute-solution interfaces, where the solute generates a strong electrostatic potential in its vicinity.\\
Therefore, the NLPBE is necessary to accurately describe ion distributions around the solute molecule. Accordingly, efficient program packages for solving the NLPBE have been developed over the past decades. DelPhi\cite{li_delphi_2012} and APBS\cite{jurrus_improvements_nodate} are among the most popular software packages, both of which employ an atomic charge representation of the solute charge distribution. While this approach produces reasonable results for huge biomolecular systems, it might be insufficient for achieving atomistic accuracy that electronic structure calculations pursue. Some NLPBE solvers support electronic structure calculations such as BigDFT,\cite{fisicaro_generalized_2016} JDFTx,\cite{letchworth-weaver_joint_2012}, and DL\_MG\cite{womack_dl_mg_2018}; however, these are primarily designed for quantum calculations under the periodic boundary conditions, which might be inappropriate for describing molecular systems. To the best of our knowledge, Q-Chem is the only software package that supports the NLPBE solver integrated with molecular calculations.\cite{shao_advances_2015,stein_poissonboltzmann_2019}\\
The limited availability of molecular NLPBE solvers motivated us to develop a library that can be easily integrated with molecular electronic structure codes. Furthermore, we aim to reduce the computational cost of solving the NLPBE, thereby expanding its applicability to a broad range of chemistry performed in electrolyte solutions.\\
Therefore, we present an open-source Python library for efficiently solving the NLPBE, namely libNLPBE. Our library features the density fitting (DF) approximation for calculating solute electrostatic potentials, the modified Damped Inexact Newton Multigrid developed by Holst (mDINMH), and GPU acceleration, all of which address the major computational bottlenecks in solving the NLPBE.\\
In the following, we first discuss the expression for the dielectric function and the size-modified ion charge density appearing in the NLPBE (Section \ref{subsec:nlpbe}). Subsequently, we introduce the finite difference method for numerically solving the NLPBE (Section \ref{subsec:discretization}). Next, we provide the theoretical background of the DF approximation for efficiently calculating solute electrostatic potentials (Section \ref{subsec:soluteESP}) and the mDINMH method for solving the NLPBE (Section \ref{subsec:mdinmh}). In Section \ref{sec:ResultsDiscussion}, we show that both the DF approximation and the mDINMH algorithm significantly accelerate solving the NLPBE for molecules. Lastly, we provide hands-on examples of using libNLPBE for electronic structure calculations (Section \ref{sec:Examples}).

\section{Theory}\label{sec:theory}
\subsection{Non-Linear Poisson-Boltzmann Equation}\label{subsec:nlpbe}
\begin{figure}
    \centering
    \includegraphics[width=6cm]{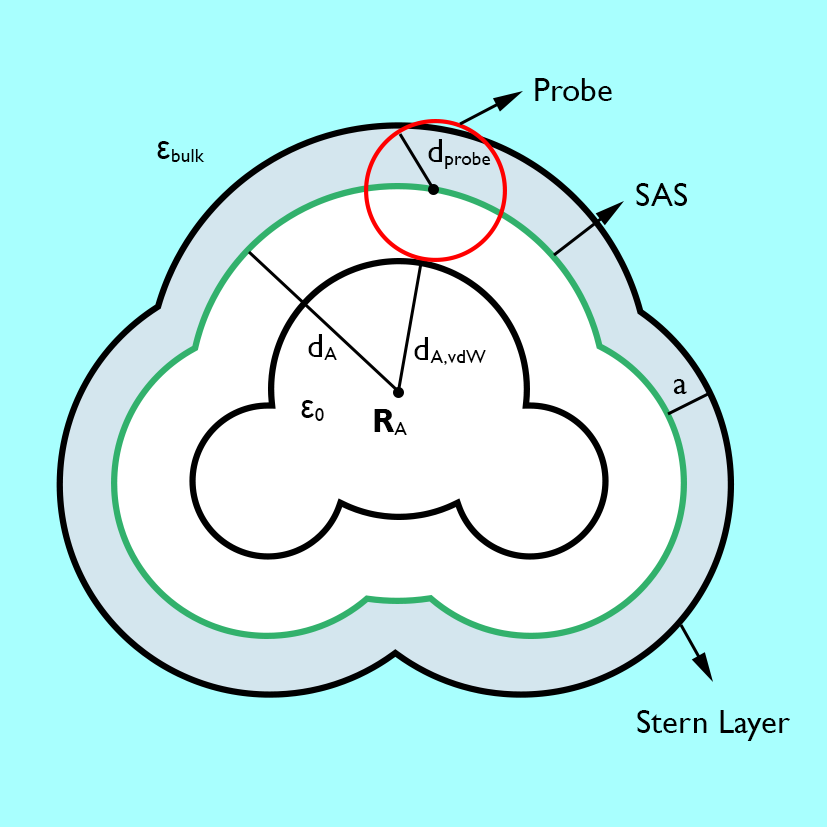}
    \caption{Schematic illustration for the solvent-accessible surface (SAS).}
    \label{SASIonExclusion}
\end{figure}

For a fixed solute charge density $\rho^\text{sol}(\mathbf{r})$ surrounded by an electrolyte solution environment, the total electrostatic potential $\phi^\text{tot}(\mathbf{r})$ is given as the solution of the non-linear Poisson-Boltzmann equation (NLPBE)\cite{sharp_calculating_1990, fisicaro_generalized_2016, stein_poissonboltzmann_2019,herbert_dielectric_2021}
\begin{equation}
    \nabla \cdot \big ( \epsilon(\mathbf{r}) \nabla \phi^\text{tot} (\mathbf{r}) \big ) = -4\pi \big ( \rho^\text{sol}(\mathbf{r}) + \rho^\text{ions}[\phi^\text{tot}] \big ),\label{NLPBE}
\end{equation}
where $\epsilon(\mathbf{r})$ and $\rho^\text{ions}[\phi^\text{tot}]$ is the dielectric function and the ion charge density, respectively. The expression for these two variables needs to be determined prior to solving the NLPBE, which we discuss in the following.\\
Determining the expression for $\epsilon(\mathbf{r})$ is not trivial, as it sensitively depends on the structure of solute-solvent interfaces. While several researchers represent $\epsilon(\mathbf{r})$ as a functional of the solute electron density $n(\mathbf{r})$,\cite{andreussi_revised_2012,ringe_function-space-based_2016} it still remains unclear whether this representation correctly captures a sudden change of the dielectric permittivity at the interface.\\
This uncertainty originates from the absence of a clear mathematical definition for solute-solvent interfaces, which leads to various expressions for $\epsilon(\mathbf{r})$ in literature.\cite{nicholls_rapid_1991,andreussi_revised_2012,fisicaro_generalized_2016} Given the ambiguity in determining solute-solvent interfaces, we adopt a rigid cavity model in expressing $\epsilon(\mathbf{r})$ for simplicity,\cite{fisicaro_generalized_2016,coons_quantum_2018,stein_poissonboltzmann_2019} where the solvent-accessible surface (SAS, $S(\mathbf{r})$) serves as the solute-solvent interface.\\
By representing the SAS as a product of atomwise functions $s_\text{A}$ ($S(\mathbf{r}) = \prod_\text{A} s_\text{A}$), $\epsilon(\mathbf{r})$ can be defined as follows (See Figure \ref{SASIonExclusion}).\cite{fisicaro_generalized_2016,coons_quantum_2018,stein_poissonboltzmann_2019}
\begin{equation}
\begin{split}
    \epsilon(\mathbf{r}) &= \epsilon_0  + \big ( \epsilon_\text{bulk} - \epsilon_0 \big ) \prod_\text{A}^\text{atoms} s_\text{A}(d_\text{A}, \Delta; |\mathbf{r} - \mathbf{R}_\text{A}|),
    \end{split}
\label{DielectricFunction}
\end{equation}
where $\epsilon_0$ and $\epsilon_\text{bulk}$ is the dielectric permittivity of vacuum and bulk solvent, respectively.\\
To ensure the smoothness of $\epsilon(\mathbf{r})$ at the interface, we represent the atomwise function $s_\text{A}$ with an error function with an interpolation parameter $\Delta$\cite{fisicaro_generalized_2016,coons_quantum_2018,stein_poissonboltzmann_2019}
\begin{equation}
    s_\text{A}(d_\text{A}, \Delta; |\mathbf{r} - \mathbf{R}_\text{A}|) =  \cfrac{1}{2} \bigg [ 1 + \text{erf} \bigg ( \cfrac{|\mathbf{r} - \mathbf{R}_\text{A}| - d_\text{A} }{\Delta}\bigg )\bigg ],
\label{atomwise}
\end{equation}
where $\mathbf{R}_\text{A}$ and $d_\text{A}$ is the position and the atom-specific length of atom A, respectively (See Figure \ref{SASIonExclusion}).\\
As a result, the structure of the SAS critically depends on the choice of $d_\text{A}$ as shown in Equation \eqref{atomwise}, because it directly determines the location of the solute-solvent interface relative to atomic positions. We generally followed the standard protocol for determining atom-specific lengths,\cite{tomasi_quantum_2005} where $d_\text{A}$ is given as the sum of the van der Waals radius and the probe radius, $d_\text{A} = d_\text{A,vdW} + d_\text{probe}$ as schematically illustrated in Figure \ref{SASIonExclusion}. Specifically, we employed the Bondi radii set for $d_\text{A,vdW}$, except for the hydrogen atom, which was set to 1.1 Å.\cite{bondi_van_1964,rowland_intermolecular_1996}\\
We next move on to the expression for ion charge density $\rho^\text{ions}(\mathbf{r})$. The key assumption in the NLPBE is Boltzmann statistics for describing ion distributions in the solution.\cite{sharp_calculating_1990} Specifically, for species $i$, its local ion concentration $c_i(\mathbf{r})$ is given as a product of its bulk concentration $c_i^\text{b}$ and the associated Boltzmann factor as below ($\beta = 1 / k_\text{B}T$, $k_\text{B}$: Boltzmann constant, $T$: absolute temperature).
\begin{equation}
    c_i(\mathbf{r}) = c_i^\text{b} \exp ( -\beta q_i e  \phi^\text{tot}(\mathbf{r}) ),
\end{equation}
where $q_i$ and $e$ is the charge of species $i$ in atomic unit and the elementary charge ($e = 1$ in atomic unit), respectively. \\
For simplicity, we consider a 1:1 electrolyte such as NaCl ($c_i^\text{b} = c^\text{b}$) throughout this work, but our theoretical framework can be expanded to multivalent electrolytes without losing generality.\cite{andrietti_exact_1976}
For a 1:1 electrolyte solution, the ion charge density is obtained from the sum of the cation ($ec_+(\mathbf{r})$) and the anion ($-ec_-(\mathbf{r})$) charge densities. The resulting $\rho^\text{ions}[\phi^\text{tot}]$ is therefore expressed as follows.
\begin{equation}
    \rho^\text{ions}[\phi^\text{tot}] = -2ec^\text{b} \sinh ( \beta e\phi^\text{tot}(\mathbf{r}) )
\end{equation}
In principle, the above expression could be provided into the NLPBE; however it is limited in incorporating some important aspects of ions. While we have assumed an implicit ion model, ions in reality have finite size. These finite-sized ions create an ion-exclusion boundary around the solute, commonly referred to as the Stern layer, which prevents ions from overlapping with the solute cavity due to steric hindrance. To account for this aspect, we introduce an ion-exclusion function $\lambda(\mathbf{r})$ in $\rho^\text{tot}[\phi^\text{tot}]$.
\begin{equation}
    \rho^\text{ions}[\phi^\text{tot}] = -2\lambda(\mathbf{r}) ec^\text{b} \sinh ( \beta e\phi^\text{tot}(\mathbf{r}) ),
\end{equation}
where $\lambda(\mathbf{r})$ is defined using a surface separated from the SAS by the Stern layer thickness $a$ as below (See Figure \ref{SASIonExclusion}). 
\begin{equation}
    \lambda(\mathbf{r}) = \prod_\text{A}^\text{atoms} \cfrac{1}{2} \bigg [ 1 + \text{erf} \bigg ( \cfrac{|\mathbf{r} - \mathbf{R}_\text{A}| - d_\text{A} - a}{\Delta} \bigg )\bigg ]
    \label{IonExclusion}
\end{equation}
\\
In addition, finite-sized ions also require modification of the Boltzmann factor due to steric effects, leading to the size-modified ion charge density as below.\cite{fisicaro_generalized_2016,ringe_function-space-based_2016,stein_poissonboltzmann_2019}
\begin{equation}
    \rho^\text{ions}[\phi^\text{tot}] = -\cfrac{2\lambda(\mathbf{r})e c^b \sinh (\beta e \phi^\text{tot}(\mathbf{r})) }{1 - \cfrac{c^b}{c_{1+2}} + \cfrac{c^b}{c_{1+2}}\lambda(\mathbf{r}) \cosh (\beta e\phi^\text{tot}(\mathbf{r}))},
\label{MPB_ion_charge}
\end{equation}
where $c_{1+2}$ indicates the maximum electrolyte concentration given by ($N_A$: Avogadro number, $p = 0.74$ for the closest-packing factor)
\begin{equation}
    c_{1+2} = \cfrac{p}{\cfrac{4}{3} \pi N_\text{A} (R_+^3 + R_-^3)},
\end{equation}
where $R_+$ and $R_-$ is the radius of the cation and the anion, respectively.\cite{stein_poissonboltzmann_2019} We used $R_+=R_-=4.3$ Å, representing the average value of the radius of hydrated Na\textsuperscript{+} and Cl\textsuperscript{–}.\cite{stein_poissonboltzmann_2019} We employ the size-modified ion charge density (Equation \eqref{MPB_ion_charge}) for solving the NLPBE, and its distribution around the solute molecule will be discussed in Section \ref{sec:ResultsDiscussion}.
\subsection{Discretization}\label{subsec:discretization}
Since the analytic solution of the NLPBE is unknown in general, numerical approaches through discretization are generally employed to obtain its numerical solution. Various discretization schemes, including finite difference,\cite{sharp_calculating_1990,nicholls_rapid_1991,stein_poissonboltzmann_2019} finite element,\cite{cortis_numerical_1997,holst_adaptive_2000} and boundary element methods\cite{geng_treecode-accelerated_2013} have been applied for numerically solving the NLPBE. Among these, we chose the finite difference method in order to leverage efficient libraries developed for solving sparse linear systems (See Section \ref{subsubsec:amg}).\\
\begin{figure}
    \centering
    \includegraphics[width=8cm]{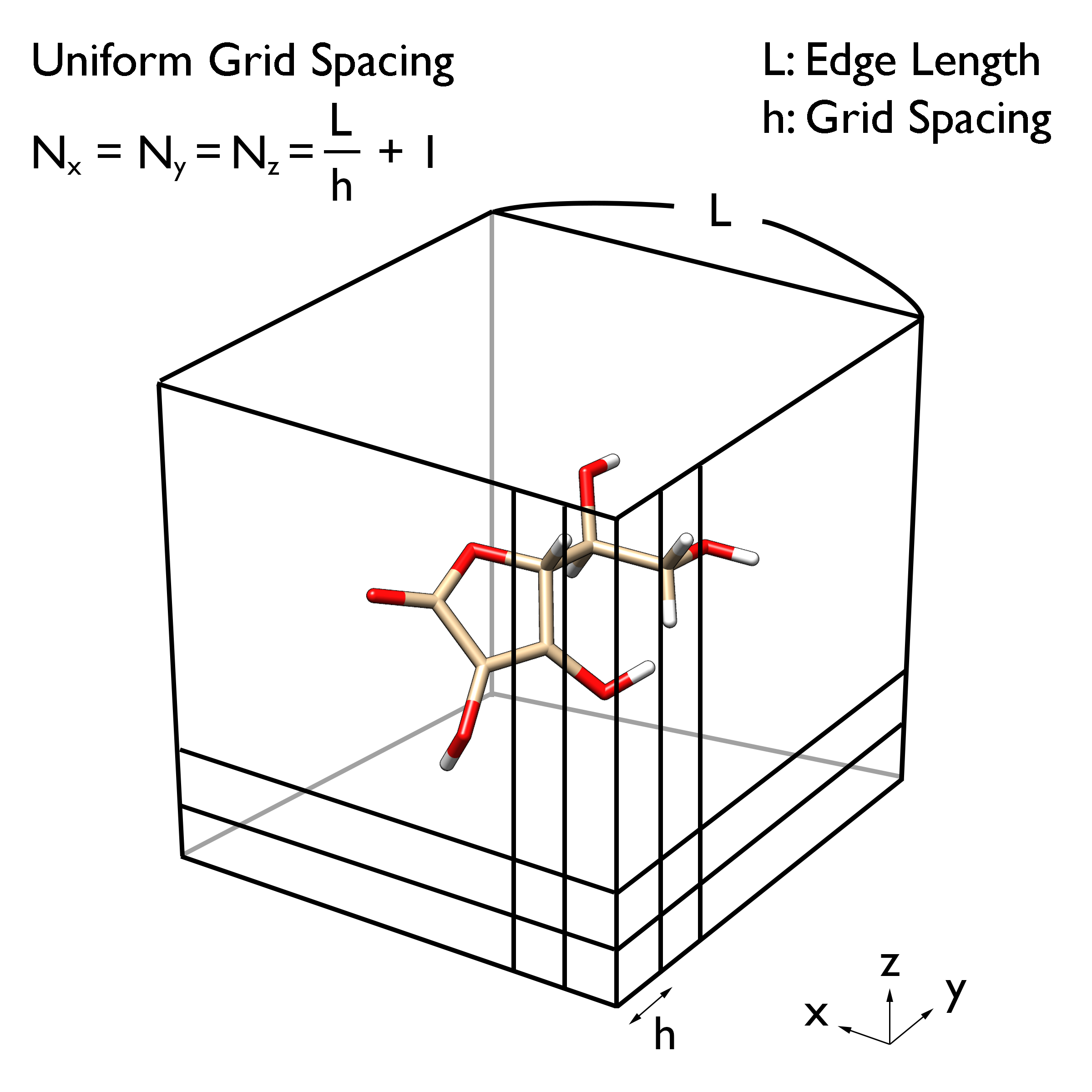}
    \caption{Schematic illustration of the finite difference method. The molecular structure of vitamin C is drawn in the Figure for illustrative purpose.}
    \label{FiniteDiff}
\end{figure}
To apply the finite difference method, we consider a cubic box with its edge length of $L$, where the solute molecule is placed at its center (Figure \ref{FiniteDiff}). Subsequently, this box is discretized into uniform Cartesian grids with $N_x$, $N_y$, and $N_z$ points with a grid spacing $h$ along each edge, resulting in a total of $N_xN_yN_z$ or $N_x^3$ grid points.\\
Similarly, the Laplacian ($\nabla^2$) and the gradient ($\nabla$) operators appearing in the NLPBE need to be discretized as well. For the Laplacian operator, we employ the 8th order central finite difference method for its discretization under the Dirichlet boundary condition, \textit{i.e.} $\phi^\text{tot}(\mathbf{r}) = 0$ at the boundaries. In contrast, the gradient operator is represented using a mixed discretization scheme: 8th order central finite difference for the interior of the cubic box and the 4th-order backward and forward finite difference method for the left-sided and the right-sided boundaries, respectively.\cite{coons_quantum_2018}

\subsection{Solute Electrostatic Potential}\label{subsec:soluteESP}
The NLPBE requires the solute charge density ($\rho^\text{sol}(\mathbf{r})$) as an input variable to be solved (Equation \eqref{NLPBE}). This quantity can be, in principle, obtained by summing the nuclear and electronic contributions to the solute charge density, $\rho^\text{sol}(\mathbf{r}) = \rho^\text{sol}_\text{nuc}(\mathbf{r}) + \rho^\text{sol}_\text{elec}(\mathbf{r})$. However, representing the nuclear charge density ($\rho^\text{sol}_\text{nuc}(\mathbf{r})$) is not trivial, as most quantum chemical calculations treat nuclei as point charges. This point charge representation leads to Coulomb singularity in $\rho^\text{sol}_\text{nuc}(\mathbf{r})$ at the nuclear positions, which poses a challenge in performing numerical calculations.\cite{coons_hydrated_2016}\\
Alternatively, $\rho^\text{sol}(\mathbf{r})$ can be indirectly obtained by solving the Poisson equation for the solute electrostatic potential ($\phi^\text{sol}(\mathbf{r})$) in vacuum.\cite{coons_quantum_2018} 
\begin{equation}
    \rho^\text{sol}(\mathbf{r}) = -\cfrac{1}{4\pi}\nabla^2 \phi^\text{sol}(\mathbf{r})
\label{PoissonEquation}
\end{equation}
This approach does not require $\rho^\text{sol}_\text{nuc}(\mathbf{r})$ to compute $\rho^\text{sol}(\mathbf{r})$, thereby avoiding the singularity issue in $\rho^\text{sol}_\text{nuc}(\mathbf{r})$. We thus adopt the indirect approach for obtaining $\rho^\text{sol}(\mathbf{r})$, and $\phi^\text{sol}(\mathbf{r})$ is the quantity that needs to be computed for solving Equation \eqref{PoissonEquation}. In the following, we discuss both the analytic and the density fitting (DF) approaches for calculating $\phi^\text{sol}(\mathbf{r})$.
\subsubsection{Analytic Approach}\label{subsubsec:analytic}
We first begin by discussing the analytic approach for computing $\phi^\text{sol}(\mathbf{r})$ at grid points.\cite{coons_quantum_2018,stein_poissonboltzmann_2019} Similar to $\rho^\text{sol}(\mathbf{r})$, $\phi^\text{sol}(\mathbf{r})$ can be split into the nuclear ($\phi^\text{sol}_\text{nuc}(\mathbf{r})$) and the electronic contributions ($\phi^\text{sol}_\text{elec}(\mathbf{r})$).
\begin{equation}
    \phi^\text{sol}(\mathbf{r}) = \phi^\text{sol}_\text{nuc}(\mathbf{r}) + \phi^\text{sol}_\text{elec}(\mathbf{r})
\end{equation}
Under the point charge representation for nuclei, $\phi^\text{sol}_\text{nuc}(\mathbf{r})$ is given as a Coulomb potential generated by nuclear charge $Z_\text{A}$ located at nuclear positions $\mathbf{R}_\text{A}$.
\begin{equation}
    \phi^\text{sol}_\text{nuc}(\mathbf{r}) = \sum_\text{A}^\text{atoms} \cfrac{Z_\text{A}}{|\mathbf{r} - \mathbf{R}_\text{A}|}
\end{equation}
Its computation only involves the calculation of the distance between grid points and the nuclear position $|\mathbf{r} - \mathbf{R}_\text{A}|$, which can be done within negligible computational time. Thus, evaluating $\phi^\text{sol}_\text{nuc}(\mathbf{r})$ does not become a computational bottleneck in obtaining $\phi^\text{sol}(\mathbf{r})$.\\
A major bottleneck arises from evaluating the electronic contribution $\phi^\text{sol}_\text{elec}(\mathbf{r})$, as we will see in the following. The electronic contribution $\phi^\text{sol}_\text{elec}(\mathbf{r})$ is expressed as a Coulomb potential generated by the solute electron density $n(\mathbf{r})$.
\begin{equation}
    \quad \phi^\text{sol}_\text{elec}(\mathbf{r}) = -\int \cfrac{n(\mathbf{r'})}{|\mathbf{r} - \mathbf{r'}|} d\mathbf{r'}
\end{equation}
We further expand $n(\mathbf{r})$ using the density matrix ($P_{\mu\nu}$) and atomic orbitals ($g_\mu(\mathbf{r})$), which leads to the following expression.
\begin{equation}
    \phi^\text{sol}_\text{elec}(\mathbf{r}) = -\sum_{\mu \nu} P_{\mu \nu} \int \cfrac{g_\mu(\mathbf{r'}) g_\nu(\mathbf{r'})}{|\mathbf{r} - \mathbf{r'}|} d\mathbf{r'}
\end{equation}
By introducing a Dirac-$\delta$ function centered at a grid point $\mathbf{r}$, the integral in the equation above can be transformed into a three-center-two-electron integral as below.
\begin{equation}
\begin{split}
    \phi^\text{sol}_\text{elec}(\mathbf{r}) &= -\sum_{\mu \nu} P_{\mu \nu} \int \cfrac{g_\mu(\mathbf{r'}) g_\nu(\mathbf{r'})}{|\mathbf{r} - \mathbf{r'}|} d\mathbf{r'} \\
    &= -\sum_{\mu \nu} P_{\mu \nu} \int \cfrac{\delta (\mathbf{r}'' - \mathbf{r}) g_\mu(\mathbf{r'}) g_\nu(\mathbf{r'})}{|\mathbf{r}'' - \mathbf{r'}|} d\mathbf{r}' d\mathbf{r}'' \\
    &= -\sum_{\mu \nu} P_{\mu \nu} (\delta|\mu \nu)
\label{3c2e}
\end{split}
\end{equation}
For practical calculations, the Dirac-$\delta$ function is approximated by a normalized Gaussian function with a sufficiently large exponent. With this representation, $\phi^\text{sol}_\text{elec}(\mathbf{r})$ can be efficiently evaluated using a robust electron repulsion integral (ERI) library, and we term this approach as the analytic approach.\\
Despite the use of highly optimized ERI libraries, the analytic approach suffers from its significant computational cost caused by the number of ERIs involved in Equation \eqref{3c2e}. For a system whose electron density is described by $N_\text{ao}$ number of atomic orbitals, the computational cost scales as $N_\text{ao}\times (N_\text{ao}+1) / 2$. This corresponds to quadratic scaling with respect to $N_\text{ao}$ ($O(N_\text{ao}^2)$) for calculating $\phi^\text{sol}_\text{elec}(\mathbf{r})$, leading to prohibitive computational cost as the system size increases. We therefore seek a more efficient approach for evaluating $\phi^\text{sol}_\text{elec}(\mathbf{r})$, which we describe in the next section.\\

\subsubsection{Density Fitting Approach}\label{subsubsec:densityfitting}
The quadratic scaling of the analytic approach motivated us to investigate a more efficient method for calculating $\phi^\text{sol}_\text{elec}(\mathbf{r})$. Inspired by the work by Zhao,\cite{zhang_density_2023} we utilize the density fitting (DF) approximation\cite{eichkorn_auxiliary_1995} to lower the quadratic scaling to a linear scaling, thereby accelerating the evaluation of $\phi^\text{sol}_\text{elec}(\mathbf{r})$.\\
The DF approach approximates the true electron density by a linear combination of auxiliary basis functions $K(\mathbf{r})$ as below.
\begin{equation}
    n(\mathbf{r}) = \sum_{\mu \nu} P_{\mu \nu} g_{\mu}(\mathbf{r}) g_\nu(\mathbf{r}) \approx \sum_K c_K K(\mathbf{r}) = n_\text{fit}(\mathbf{r})
\end{equation}
The fitting coefficients $c_K$ are determined by minimizing the Coulomb metric error $\mathcal{L}[\{ c_K\}]$ defined as\cite{eichkorn_auxiliary_1995} 
\begin{equation}
    \mathcal{L}[\{c_K\}] = \int \cfrac{[n(\mathbf{r}) - n_\text{fit}(\mathbf{r})][n(\mathbf{r'}) - n_\text{fit}(\mathbf{r'})]}{|\mathbf{r} - \mathbf{r'}|} d\mathbf{r} d\mathbf{r'}.
\end{equation}
Setting $\partial \mathcal{L}/\partial c_K = 0$ gives the standard resolution of the identity equation for Coulomb potentials
\begin{equation}
    \sum_{\mu \nu} P_{\mu \nu} (\mu \nu | L) = \sum_K c_K (K|L),
\end{equation}
which can be efficiently solved for $c_K$ using the Cholesky decomposition.\\
Similar to the analytic approach (Equation \eqref{3c2e}), $\phi^\text{sol}_\text{elec}(\mathbf{r})$ under the DF approximation, which we term the DF approach, can be evaluated by introducing a Dirac-$\delta$ function centered at a grid point $\mathbf{r}$, which reformulates $\phi^\text{sol}_\text{elec}(\mathbf{r})$ in terms of two-center-two-electron integrals as below.
\begin{equation}
\begin{split}
    \phi^\text{sol}_\text{elec}(\mathbf{r}) &\approx -\sum_K c_K \int \cfrac{\delta (\mathbf{r}'' - \mathbf{r} )K(\mathbf{r'})}{|\mathbf{r}'' - \mathbf{r}'|} d\mathbf{r}'d\mathbf{r}''  \\
    &= -\sum_{K} c_K (\delta | K )
\label{2c2e}
\end{split}
\end{equation}
Thus, the total number of ERIs involved in Equation \eqref{2c2e} scales to the number of auxiliary basis functions ($N_\text{aux}$), indicating linear computational scaling with respect to $N_\text{aux}$ ($O(N_\text{aux})$) for computing $\phi^\text{sol}_\text{elec}(\mathbf{r})$. Furthermore, given that $N_\text{aux} \ll N_\text{ao}\times (N_\text{ao}+ 1)/2$ in most cases, we expect that the DF approach will show dramatic acceleration compared to the analytic approach, as will be demonstrated in Section \ref{sec:ResultsDiscussion}.\\

\subsection{Modified Damped Inexact Newton Method}\label{subsec:mdinmh}
This section discusses the modified Damped Inexact Newton Multigrid developed by Holst (mDINMH) method\cite{holst_numerical_1995} as an efficient algorithm for solving the NLPBE. The complete algorithm is outlined in Algorithm \ref{DINMHAlgorithm}, and readers are encouraged to refer to it while following the discussion in this section.\\
\subsubsection{Inexact Newton Method}\label{subsubsec:inexactnewton}
We begin with introducing the inexact Newton method for solving the NLPBE. Define a functional $\mathcal{G}[\phi^\text{tot}]$ by rearranging Equation \eqref{NLPBE}
\begin{equation}
    \mathcal{G}[\phi^\text{tot}]  = -\mathcal{A}\phi^\text{tot}(\mathbf{r}) - \cfrac{4\pi}{\epsilon(\mathbf{r})} \big ( \rho^\text{sol}(\mathbf{r}) + \rho^\text{ions}[\phi^\text{tot}]\big ),
\end{equation}
 where the operator $\mathcal{A}$ is given as 
 \begin{equation}
     \mathcal{A} = \nabla \ln \epsilon(\mathbf{r}) \cdot \nabla + \nabla^2.
 \label{OperatorA}
 \end{equation}
Note that our functional $\mathcal{G}[\phi^\text{tot}]$ differs from that in the original literature,\cite{holst_numerical_1995} which defines $\mathcal{F}[\phi^\text{tot}] = -\nabla \cdot ( \epsilon(\mathbf{r}) \nabla \phi^\text{tot}(\mathbf{r}) ) - 4\pi ( \rho^\text{sol}(\mathbf{r}) + \rho^\text{ions}[\phi^\text{tot}] )$. While this modification is made for implementation convenience, $\mathcal{G}[\phi^\text{tot}]$ still retains the important mathematical properties of $\mathcal{F}[\phi^\text{tot}]$ used for solving the NLPBE.\\
By defining $\mathcal{G}[\phi^\text{tot}]$, solving the NLPBE becomes equivalent to finding its root. Given that its Jacobian $\mathcal{G}'[\phi^\text{tot}]$ can be easily found, it is straightforward to apply the Newton method for locating the root of $\mathcal{G}[\phi^\text{tot}]$. Thus, we aim to iteratively update the solution  $\phi^\text{tot}_{k+1} = \phi^\text{tot}_k + v^k$ through the Newton method until $||\mathcal{G}[\phi^\text{tot}_k]||$ becomes sufficiently close to zero (\textit{for} loop for $k$ in Algorithm \ref{DINMHAlgorithm}, black arrows in Figure S1).\\
The update term $v^k$ is obtained by solving the following Newton equation.
\begin{equation}
    \mathcal{G}'[\phi^\text{tot}_k]v^k = -\mathcal{G}[\phi^\text{tot}_k],
\label{ExactNewton}
\end{equation}
where the Jacobian $\mathcal{G}'[\phi^\text{tot}_k]$ is
\begin{equation}
    \mathcal{G}'[\phi^\text{tot}_k] = -\mathcal{A} - \cfrac{4\pi}{\epsilon(\mathbf{r})} \cdot \cfrac{\delta \rho^\text{ions}}{\delta \phi^\text{tot}} \bigg |_{\phi^\text{tot}_k}.
\label{Jacobian}
\end{equation}
Equation \eqref{ExactNewton} also requires an iterative method to find $v^k$, which updates a guess solution $v^k_i$ through $v^k_{i+1} = v^k_i + \Delta v^k_i$ until the residual $r^k_i = \mathcal{G}'[\phi^\text{tot}_k] v_i^k + \mathcal{G}[\phi^\text{tot}_k]$ gets sufficiently close to zero.\\
These steps constitute the exact Newton method and should be applicable to solving the NLPBE in principle. However, solving Equation \eqref{ExactNewton} at every Newton step leads to significant computational effort. Alternatively, it is sufficient to solve the Newton equation inexactly, provided that the current $v_i^k$ is guaranteed to be a descent direction, \textit{i.e.} $||\mathcal{G}[\phi^\text{tot}_{k+1}]|| < ||\mathcal{G}[\phi^\text{tot}_{k}] ||$. This approach substantially reduces computational time, which is the motivation of the inexact Newton method in the DINMH, which we describe in the following.\cite{holst_numerical_1995}\\
We rewrite Equation \eqref{ExactNewton} with the residual $r^k_i$ to ensure that the Newton equation is solved inexactly.
\begin{equation}
    \mathcal{G}'[\phi^\text{tot}_k] v_i^k = -\mathcal{G}[\phi^\text{tot}_k] + r_i^k,
\label{InexactNewton}
\end{equation}
and we stop iterations for index $i$ (\textit{for} loop for $i$ in Algorithm \ref{DINMHAlgorithm}, red arrows in Figure S1) once the current $v^k_i$ is assured to be a descent direction. According to Holst,\cite{holst_numerical_1995} a descent $v_i^k$ is guaranteed if $r_i^k$ satisfies the following condition.
\begin{equation}
    ||r_i^k|| < ||\mathcal{G}[\phi^\text{tot}_k] ||.
\label{DescentCondition}
\end{equation}
Additionally, we ensure local $Q$-order($1+p$) convergence of the inexact Newton method by enforcing the following condition.\cite{holst_numerical_1995}
\begin{equation}
    ||r_i^k|| \leq C||\mathcal{G}[\phi^\text{tot}_k]||^{1+p}
\label{Qorder1+p}
\end{equation}
This condition allows the algorithm to focus more on finding a descent direction through Equation \eqref{InexactNewton}, rather than achieving high accuracy when $\phi^\text{tot}_k$ is far from the solution $\phi^\text{tot}$. Conversely, it demands greater accuracy when solving Equation \eqref{InexactNewton} as $\phi^\text{tot}_k$ approaches $\phi^\text{tot}$. Thus, Equation \eqref{Qorder1+p} dynamically controls the convergence threshold for solving Equation \eqref{InexactNewton}, thereby reducing the overall computational effort.\cite{dembo_inexact_1982,holst_numerical_1995}\\
Taken together, we iteratively update $v_i^k$ by $v^k_{i+1} = v^k_i + \Delta v^k_i$ until the residual $r_i^k$ satisfies both the descent direction (Equation \eqref{DescentCondition}) and $Q$-order($1+p$) convergence (Equation \eqref{Qorder1+p}) conditions as summarized in Algorithm \ref{DINMHAlgorithm} (\textit{for} loop for $i$ in Algorithm \ref{DINMHAlgorithm}, red arrows in Figure S1). The update $\Delta v_i^k$ at $i$-th step is obtained by assuming that the residual at the next step vanishes ($r^k_{i+1} \rightarrow 0$), which leads to the following equation.
\begin{equation}
    \mathcal{G}'[\phi^\text{tot}_k] \Delta v_i^k = -r_i^k
\label{InexactNewtonInnerDelta}
\end{equation}
This is the key equation of the inexact Newton method, and the detailed algorithm for solving it is discussed in the next section.
\begin{algorithm}
\caption{Algorithm of the modified DINMH method. Its flowchart is also provided Figure S1 in Supplementary Information.}
Initialize $\mathcal{P}$ and multigrid hierarchy\\
$\phi^\text{tot}_0 = 0$\\
\For{$k = 0$}{
    $\mathcal{G}[\phi^\text{tot}_k]$, $\mathcal{G}'[\phi^\text{tot}_k]$\\
    \If{$||\mathcal{G}[\phi^\text{tot}_k]|| < \text{threshold}$} {
        $\phi^\text{tot} = \phi^\text{tot}_k$\\
        \textbf{return} $\phi^\text{tot}$\\
    }
    $v_0^k = 0$ \\
    \For{$i = 0$} {
        $r_i^k = \mathcal{G}'[\phi^\text{tot}_k]v_i^k + \mathcal{G}[\phi^\text{tot}_k]$ \\
        \If{$||r_i^k|| < ||\mathcal{G}[\phi^\text{tot}_k] ||$ and $||r_i^k|| \leq C||\mathcal{G}[\phi^\text{tot}_k] ||^{p+1}$} {
        $v^k \leftarrow v^k_i$\\
        \textbf{break}\\
        }
       $\mathcal{P}\Delta v_i^k = -r_i^k$\\
        $v^k_{i+1} \leftarrow v_i^k + \Delta v_i^k$\\
    }
    $\lambda_k^0 = 1$ \\
    \For{$n = 0$} {
        \If{$||\mathcal{G}[\phi^\text{tot}_k + \lambda_k^n v^k] || < || \mathcal{G}[\phi^\text{tot}_k] ||$} {
        $\lambda_k = \lambda_k^n$\\
        \textbf{break}
        }
        $\lambda_k^{n+1} = \lambda_k^n / 2$
        }
    $\phi^\text{tot}_{k+1} = \phi^\text{tot}_k + \lambda_k v^k$\\
}
\label{DINMHAlgorithm}
\end{algorithm}
\subsubsection{Preconditioning}\label{subsubsec:precond}
The goal of the inexact Newton method aligns with solving Equation \eqref{InexactNewtonInnerDelta}. In the original DINMH, this equation is solved using a multigrid method enabled by a symmetrized representation for $\nabla \epsilon(\mathbf{r}) \cdot \nabla$ through discretization.\cite{holst_numerical_1995,womack_dl_mg_2018} Such a symmetrization is essential for most multigrid methods because, without it, the Jacobian is asymmetric in general and may require a more sophisticated method, such as the generalized minimal residual (GMRES) method.\\
These considerations drive us to introduce a symmetric preconditioner $\mathcal{P}$ by dropping asymmetric $\nabla \ln \epsilon(\mathbf{r}) \cdot \nabla$ term in $\mathcal{G}'[\phi^\text{tot}]$ as below.
\begin{equation}
    \mathcal{P} = -\nabla^2 - \cfrac{4\pi}{\epsilon(\mathbf{r})} \cdot \cfrac{\delta \rho^\text{ions}}{\delta \phi^\text{tot}} \bigg |_{0}
\label{Preconditioner}
\end{equation}
Moreover, we use a constant $\delta \rho^\text{ions}/\delta \phi^\text{tot}$ evaluated at $\phi^\text{tot} = 0$ in order to fix $\mathcal{P}$ throughout the entire iterative procedure, including the electronic SCF cycles. This is intended to avoid reconstruction of the multigrid hierarchy, which we will discuss in the next section (Section \ref{subsubsec:amg}).\\
Thus, instead of directly solving Equation \eqref{InexactNewtonInnerDelta}, we solve its preconditioned form to find $\Delta v_i^k$.
\begin{equation}
    \mathcal{P} \Delta v_i^k = -r_i^k
\label{InexactNewtonInnerDeltaPrecond}
\end{equation}
As $\mathcal{P}$ is a symmetric operator by construction, Equation \eqref{InexactNewtonInnerDeltaPrecond} can be solved using an efficient multigrid method developed for symmetric operators. Specifically, we perform the preconditioned defect correction method, where a single V-cycle sweep serves as an approximate inverse of $\mathcal{P}$. Subsequently, the resulting $\Delta v_i^k$ is added to $v_i^k$ to obtain $v_i^{k+1}$ ($v_{i+1}^k = v_i^k + \Delta v_i^k$). We then iterate these steps until the updated residual $r_{i+1}^k$ satisfies both the descent direction (Equation \eqref{DescentCondition}) and the $Q$-order($1+p$) convergence (Equation \eqref{Qorder1+p}) conditions.\\
Compared to the original DINMH method, our approach, which we term the modified DINMH (mDINMH), features the introduction of a symmetric preconditioner (Equation \eqref{Preconditioner}). As a result, the mDINMH method does not require a symmetric representation for $\nabla \ln \epsilon(\mathbf{r}) \cdot \nabla$ through discretization, thereby allowing $\nabla \ln \epsilon(\mathbf{r})$ and $\nabla$ to be treated separately. This feature enables the use of an analytic expression for $\nabla \ln \epsilon(\mathbf{r})$ if available, which is expected to reduce discretization errors.

\subsubsection{Algebraic Multigrid Method}\label{subsubsec:amg}
\begin{figure}
    \centering
    \includegraphics[width=8cm]{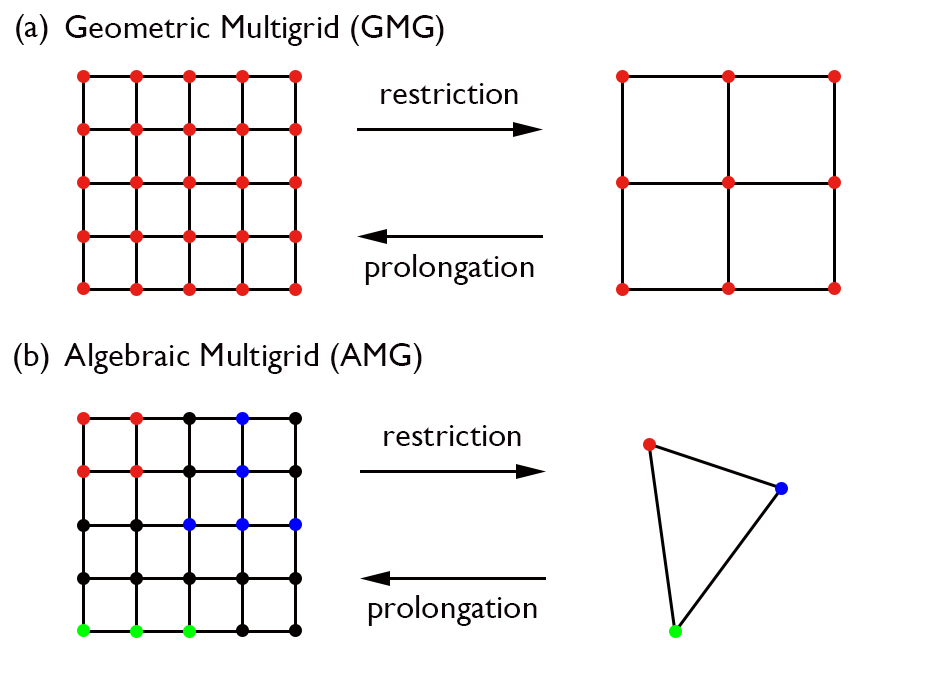}
    \caption{Schematic illustration of (a) geometric multigrid (GMG) and (b) algebraic multigrid (AMG).}
    \label{GMGvsAMG}
\end{figure}

Our preconditioner $\mathcal{P}$ (Equation \eqref{Preconditioner}) uses a constant $\delta \rho^\text{ions}/\delta \phi^\text{tot}$ evaluated at $\phi^\text{tot} = 0$. This is to overcome limitations that the original DINMH method suffers from.\cite{holst_numerical_1995,womack_dl_mg_2018}\\
In the original work,\cite{holst_numerical_1995} Equation \eqref{InexactNewtonInnerDelta} is solved using a geometric multigrid (GMG) method. GMG exploits a multigrid hierarchy automatically constructed from the grid structure ($2^n + 1 \rightarrow 2^{n-1}+1 \rightarrow \cdots$, See Figure \ref{GMGvsAMG}a), along with the corresponding restriction and prolongation operators. While this feature allows GMG methods to avoid explicit construction of multigrid hierarchies, it results in strict restrictions on the number of grid points along each spatial direction ($x$,$y$, and $z$-axis) to achieve optimal efficiency as a trade-off for this advantage.\\
To relax these restrictions while maintaining efficiency, we adopt algebraic multigrid (AMG) as a black-box method for solving Equation \eqref{InexactNewtonInnerDeltaPrecond} (See Computationl Details for its settings). As AMG algebraically determines multigrid hierarchies, it no longer imposes any restrictions on the number of grid points for optimal efficiency. This feature provides greater flexibility in the choice grid points than GMG, making AMG be applicable for general purpose.\\
Despite the flexibility of AMG, its major drawback arises from the construction of its multigrid hierarchy. As schematically shown in Figure \ref{GMGvsAMG}b, the hierarchy depends on the structure of the underlying linear operator. This indicates that the hierarchy has to be reconstructed whenever the operator changes, which results in significant computational overhead.\\
Fortunately, this overhead can be minimized by using a fixed linear operator, which motivates the use of a constant preconditioner as defined in Equation \eqref{Preconditioner}. By setting a constant $\delta \rho^\text{ions}/\delta \phi^\text{tot}$ in $\mathcal{P}$, the AMG hierarchy remains unchanged throughout the entire calculations, thereby allowing its construction only once.\\
Therefore, the use of AMG alleviates the restrictions on grid points that GMG in the original DINMH suffers from. Furthermore, combined with a constant preconditioner, AMG can minimize the computational overhead associated with reconstruction of the AMG hierarchy, thereby offering both flexibility and computational efficiency. 

\subsubsection{Damping}\label{subsubsec:damping}
The DINMH method also employs a damping factor ($\lambda_k$) when updating $\phi^\text{tot}_k$ ($\phi^\text{tot}_{k+1} = \phi^\text{tot}_k + \lambda_k v^k$) to ensure global convergence of the inexact Newton method (Algorithm \ref{DINMHAlgorithm}, blue arrows in Figure S1).\cite{holst_numerical_1995} Specifically, once a descent direction $v^k$ is found, a damping factor $\lambda_k$ is determined by the backtracking line search with its initial value of 1 ($\lambda_k^0 = 1$) until the following condition is satisfied. 
\begin{equation}
    || \mathcal{G}[\phi^\text{tot}_k + \lambda_k^n v^k] || < || \mathcal{G}[\phi^\text{tot}_k] ||
\end{equation}
If the condition is not satisfied, the damping factor is updated by halving the current value, \textit{i.e.} $\lambda_k^{n+1} \leftarrow \lambda_k^n / 2$, and the condition is reexamined.\\
The entire algorithm of the mDINMH is given in Algorithm \ref{DINMHAlgorithm}, and its performance will be demonstrated in Section \ref{sec:ResultsDiscussion}.

\subsection{GPU Acceleration}\label{subsubsec:gpu}
As seen in Section \ref{subsec:soluteESP}, both the analytic and the DF approaches for $\phi^\text{sol}_\text{elec}(\mathbf{r})$ require the evaluation of ERIs for all grid points. This results in a computational cost scaling as $O(N_x^3)$, where $N_x$ is the number of grid points along $x$, $y$, and $z$-axis (See Figure \ref{FiniteDiff}). This cubic growth rapidly makes $\phi^\text{sol}_\text{elec}(\mathbf{r})$ computationally intractable as the number of grid points increases. For example, setting $N_x = 97$ produces 912,673 grid points, indicating that the total number of ERIs for $\phi^\text{sol}_\text{elec}(\mathbf{r})$ is equal to $N_\text{ao}\times (N_\text{ao}+1) / 2 \times 912,673$ and $N_\text{aux} \times 912,673$ for the analytic and the density fitting approach, respectively.  Even for a few hundred $N_\text{ao}$ and $N_\text{aux}$, the number of ERIs easily exceeds ten millions due to the cubic scaling with respect to $N_x$. This gigantic number of ERIs significantly increases the computational cost of $\phi^\text{sol}_\text{elec}(\mathbf{r})$ calculations, requiring massive parallelism to obtain $\phi^\text{sol}_\text{elec}(\mathbf{r})$ within a reasonable time.\\
To mitigate this computational bottleneck, we leverage the parallelization efficiency of GPU computations. Fortunately, GPU4PySCF offers a GPU-accelerated ERI library,\cite{li_introducing_2025} and we adapt it to accelerate $\phi^\text{sol}_\text{elec}(\mathbf{r})$ calculations for both the analytic and the DF approaches.\\
Furthermore, the mDINMH can be GPU-accelerated by integrating AMGCL library, a header-only C++ AMG library for both CPU and GPU.\cite{demidov_amgcl_2019,demidov_amgcl_2020,demidov_accelerating_2021} Thus, we implemented a wrapper that allows the AMGCL functions to be accessed at the Python level. \\
To fully exploit GPU acceleration and avoid the overhead associated with CPU-GPU data transfers, we release a separate CuPy-based library, namely GPU4libNLPBE, as a GPU-accelerated version of libNLPBE. The computational efficiency of GPU4libNLPBE will also be discussed in Section \ref{sec:ResultsDiscussion}.

\section{Computational Details}\label{sec:ComputationalDetails}
All density functional calculations\cite{hohenberg_inhomogeneous_1964,kohn_self-consistent_1965,parr_density-functional_1995} were carried out at B3LYP-D3/def2-TZVPPD level of theory\cite{becke_densityfunctional_1993,lee_development_1988, vosko_accurate_1980, becke_density-functional_1988,grimme_consistent_2010,weigend_balanced_2005} as implemented in PySCF.\cite{sun_recent_2020} Resolution of the identity approximation with def2-TZVPP-JKFIT auxiliary basis set\cite{weigend_ri-mp2_1998} was generally adopted to accelerate density functional calculations. The analytic and the DF approaches for $\phi^\text{sol}_\text{elec}(\mathbf{r})$ were implemented using PySCF\cite{sun_recent_2020} and GPU4PySCF\cite{li_introducing_2025} for CPU and GPU version of libNLPBE, respectively. The algebraic multigrid method in the mDINMH method was implemented using AMGCL library\cite{demidov_amgcl_2019,demidov_amgcl_2020,demidov_accelerating_2021} for both CPU and GPU version of libNLPBE. After extensive numerical experiments, we found that 4-level V-cycle multigrid with the sparse approximate inverse relaxation of the zeroth variant (SPAI-0) relaxation method\cite{broker_sparse_2002} delivers optimal efficiency, During travering the multigrid hierarchy, pre- and post-smoothening were performed once at each level, and the coarsest level was directly solved.\\
For the dielectric and the ion-exclusion functions (Equation \eqref{DielectricFunction} and \eqref{IonExclusion}), we used $\Delta = 0.265$ Å, $a = 0.44$ Å, and $d_\text{probe} = 1.4$ Å according to the literature, \cite{stein_poissonboltzmann_2019} and the room temperature ($T = 298.15$ K) is generally considered throughout this work.
\\
We enforced the convergence threshold of $||\mathcal{G}[\phi^\text{tot}_k]|| < 1.0 \times 10^{-9}$ for the mDINMH. To demonstrate the performance of the mDINMH, we used the self-consistent (SC) method as the reference method,\cite{fisicaro_generalized_2016,coons_hydrated_2016,coons_quantum_2018,stein_poissonboltzmann_2019} and, to the best of our knowledge, it is the sole method that has been applied to molecular calculations.\cite{stein_poissonboltzmann_2019,kim_fractional_2026} Its algorithm\cite{fisicaro_generalized_2016,coons_hydrated_2016,coons_quantum_2018,stein_poissonboltzmann_2019} and implementation details\cite{kim_fractional_2026} can be found elsewhere.\\
We note that the SC method was also implemented using AMGCL library, which allows rigorous performance comparison between the SC and the mDINMH methods. We also note that the convergence scheme for the SC method is different from that of mDINMH. To be specific, the SC method uses the ion charge density $\rho^\text{ions}(\mathbf{r})$ and the polarization charge density $\rho^\text{pol}(\mathbf{r})$ for convergence test, where the latter is defined as ($\rho^\text{tot}(\mathbf{r}) = \rho^\text{sol}(\mathbf{r}) + \rho^\text{ions}(\mathbf{r})$)
\begin{equation}
    \rho^\text{pol}(\mathbf{r}) = \cfrac{1 - \epsilon(\mathbf{r})}{\epsilon(\mathbf{r})} \rho^\text{tot}(\mathbf{r}) + \cfrac{1}{4\pi} \nabla \ln \epsilon(\mathbf{r}) \cdot \nabla \phi^\text{tot}(\mathbf{r}).
\label{rho_pol}
\end{equation}
Convergence of the SC method is examined by testing whether $\Delta \rho^\text{ions}(\mathbf{r})$ and $\Delta \rho^\text{pol}(\mathbf{r})$ are smaller than their respective threshold value for all grid points. We set $\Delta\rho^\text{pol}(\mathbf{r})=1.0\times10^{-5}$ a.u. and $\Delta \rho^\text{ions}(\mathbf{r})=1.0\times10^{-6}$ a.u. for these threshold values, which are two orders of magnitude tighter than those in the original work.\cite{stein_poissonboltzmann_2019} Accordingly, the convergence threshold for the conjugate gradient step on each multigrid level was set to $1.0\times10^{-8}$ a.u.. The reference data were calculated by Q-Chem 5.4.\cite{shao_advances_2015} For consistency with PySCF-calculated DFT results, 75 radial and 302 angular grids for numerical integrations and 50 radial and 194 angular grids for non-linear correlations. All computational results reported in this paper were produced using 8 threads of Intel Xeon Gold 6240 CPU, a NVIDIA GeForce RTX 2080 Ti 11GB GPU card, and 10000 MB of memory.\\

\section{Results and Discussion}\label{sec:ResultsDiscussion}
\begin{table}
\caption{Comparison between the analytic and the density fitting (DF) approach for calculating the electrostatic potential of vitamin C and 4-nitroaniline. Calculations were performed with a 20 Å cubic box discretized into 97$\times$97$\times$97 cubic grids. The solute electron density is computed at B3LYP-D3/def2-TZVPPD level of theory.}
\begin{tabular}{ccccccc}
\hline \hline
 & \multicolumn{3}{c}{Vitamin C} & \multicolumn{3}{c}{4-nitroaniline} \\
\hline
 & analytic & DF & speedup & analytic & DF & speedup \\
\hline
CPU (sec) & 1164.88 & 14.21 &  81.98 & 750.23 & 10.97 &  68.39 \\
GPU (sec) & 38.11   & 3.86  &  9.87  & 23.68  & 3.74  &  6.33  \\              
\hline
$N_\text{ao}$  & \multicolumn{3}{c}{598}  & \multicolumn{3}{c}{478} \\
$N_\text{aux}$ & \multicolumn{3}{c}{1056} & \multicolumn{3}{c}{866} \\
$\frac{N_\text{ao}(N_\text{ao}+1)}{2N_\text{aux}}$ & \multicolumn{3}{c}{169.60} & \multicolumn{3}{c}{132.20} \\
\hline \hline
\end{tabular}
\label{AnalyticvsDensityFitting}
\end{table}

\begin{figure}
    \centering
    \includegraphics[width=8cm]{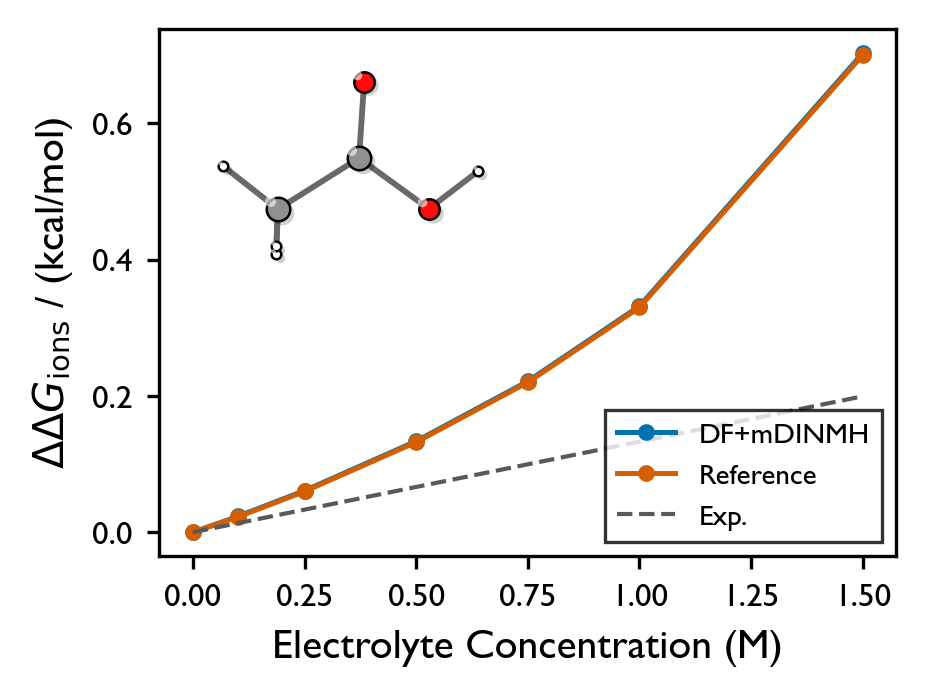}
    \caption{Calculated $\Delta\Delta G_\text{ions}$ (in kcal/mol) of acetic acid with varying the concentration of NaCl aqueous solution. Calculations were performed with a 20 Å cubic box discretized into 97$\times$97$\times$97 cubic grids. The solute electron density is computed at $\omega$B97X-V/def2-TZVPP level of theory.\cite{mardirossian_b97x-v_2014}}
    \label{AceticAcid}
\end{figure}

\textbf{Density Fitting.} Table \ref{AnalyticvsDensityFitting} summarizes the computational time for calculating $\phi^\text{sol}_\text{elec}(\mathbf{r})$ of vitamin C and 4-nitroaniline using the analytic and the density fitting (DF) approaches on CPU and GPU for 97$\times$97$\times$97 cubic grids. The DF approach on CPU shows dramatic speedups of $\times$81.98 and $\times$68.39 for vitamin C and 4-nitroaniline, respectively, both of which highlight the efficiency of the DF approach.\\
These accelerations are enabled by a reduced number of ERIs involved in $\phi^\text{sol}_\text{elec}(\mathbf{r})$ as described in Section \ref{subsec:soluteESP} (See Equation \eqref{3c2e} and \eqref{2c2e}). Comparing the ratio of the numbers of ERIs required by the analytic and the DF approaches (See Table \ref{AnalyticvsDensityFitting}), the theoretical maximum speedup is estimated to be $\times$169.60 and $\times$132.20 for vitamin C and 4-nitroaniline, respectively. The actual speedup is less than the theoretical estimate because the entire calculation is divided into smaller batches to reduce the memory cost, resulting in additional computational overhead.\\ This overhead becomes more noticeable when the calculations are performed on GPU (See Table \ref{AnalyticvsDensityFitting}), where the actual speedup is $\times$9.87 and $\times$6.33 for vitamin C and 4-nitroaniline, respectively. The lower speedups observed on GPUs are expected because the analytic approach benefits more from GPU parallelization than the density fitting approach due to a more ERIs for evaluating $\phi^\text{sol}_\text{elec}(\mathbf{r})$.\\
For vitamin C and 4-nitroaniline, the CPU-based analytic approach required 1164.88 and 750.23 seconds for $\phi^\text{sol}_\text{elec}(\mathbf{r})$, whereas the GPU-accelerated DF approach took only 3.86 and 3.74 seconds. These numbers suggest $\times$301.78 and $\times$200.60 acceleration for vitamin C and 4-nitroaniline, confirming the efficiency of both the DF approach and GPU acceleration for computing $\phi^\text{sol}_\text{elec}(\mathbf{r})$ on grids.\\
\begin{table}
\caption{Comparison between the self-consistent (SC) and the modified DINMH (mDINMH) for solving the non-linear Poisson-Boltzmann equation of vitamin C and 4-nitroaniline in 1 M NaCl aqueous solution. Calculations were performed with a 20 Å cubic box discretized into 97$\times$97$\times$97 cubic grids. The solute electron density is computed at B3LYP-D3/def2-TZVPPD level of theory.}
\begin{tabular}{ccccccc}
\hline \hline
    & \multicolumn{3}{c}{Vitamin C} & \multicolumn{3}{c}{4-nitroaniline} \\
\hline
 & SC & mDINMH & speedup & SC & mDINMH & speedup \\
\hline
CPU (sec) & 41.84 & 10.03 & 4.17 & 43.87 & 10.37 & 4.23  \\
GPU (sec) &   –   &  0.57 &  –   &   –   & 0.59  &   –   \\
\hline \hline
\end{tabular}
\label{SCvsmDINMHtime}
\end{table}
\\
\textbf{mDINMN.} We next evaluate the performance of the mDINMH method by comparing it with the self-consistent (SC) method.\cite{fisicaro_generalized_2016,coons_hydrated_2016,coons_quantum_2018,stein_poissonboltzmann_2019}  Vitamin C and 4-nitroaniline were again chosen as representative examples, and Table \ref{SCvsmDINMHtime} summarizes the computational times required to solve the NLPBE on CPU and GPU.\\
Overall, the mDINMH (10.03 and 10.37 seconds for vitamin C and 4-nitroaniline) outperforms the SC approach (41.84 and 43.87 seconds for vitamin C and 4-nitroaniline) for both molecules in terms of computing time, approximately a 4-fold speedup on the CPU. This speedup clearly demonstrates the efficiency of the mDINMH over the SC approach.\\
Furthermore, the GPU-accelerated mDINMH shows a remarkable computing time, which only takes 0.57 and 0.59 seconds for vitamin C and 4-nitroaniline, respectively (Table \ref{SCvsmDINMHtime}). Compared to the SC method on CPU, these results correspond to $\times$73.40 and $\times$74.36 acceleration for vitamin C and 4-nitroaniline, respectively, highlighting the efficiency of GPU parallelization. While we intentionally omitted implementing the GPU-accelerated version of the SC approach due to the efficiency of the mDINMN on CPU, we do not expect that it would outperform the GPU-accelerated mDINMH, given the performance gap observed on CPU.\\
\begin{figure}
    \centering
    \includegraphics[width=8cm]{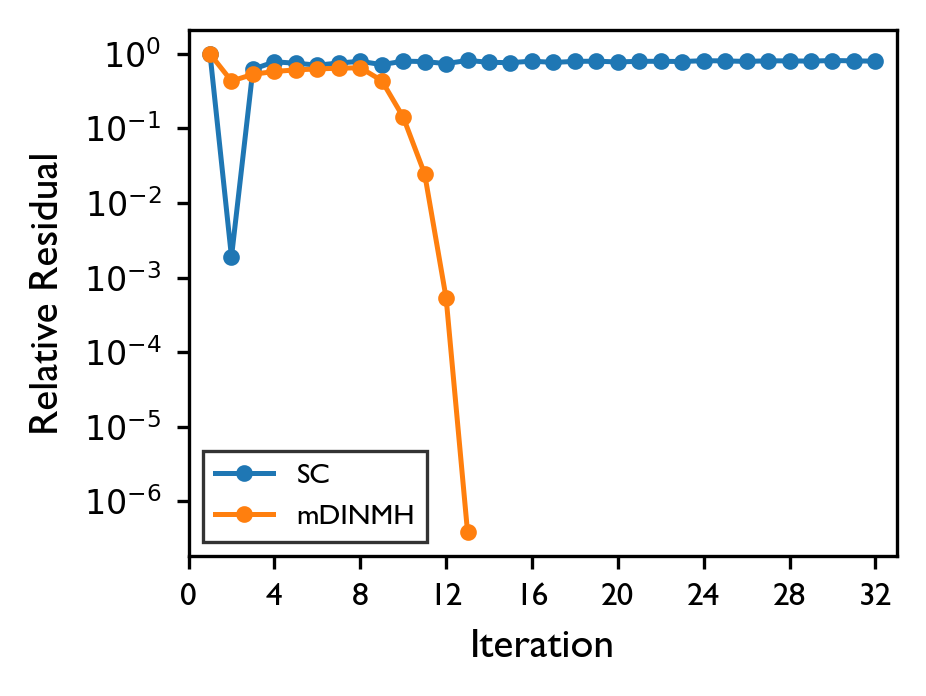}
    \caption{Relative residual vs. iteration count for solving the NLPBE of vitamin C in 1 M NaCl aqueous solution using the self-consistent (SC) (blue) and the mDINMH (orange) methods. Calculations were performed with a 20 Å cubic box discretized into 97$\times$97$\times$97 cubic grids. The solute electron density is computed at B3LYP-D3/def2-TZVPPD level of theory.}
    \label{SCvsmDINMHConvergenceRate}
\end{figure}
\\
\textbf{Convergence Rate.} The superior efficiency of the mDINMH over the SC method is primarily attributed to its faster convergence rate. To investigate the convergence rate, we plotted the relative residual ($||\mathcal{G}[\phi^\text{tot}_{k+1}]|| / ||\mathcal{G}[\phi^\text{tot}_k]||$) of both methods according to their iterative steps as illustrated in Figure \ref{SCvsmDINMHConvergenceRate}.\\
The relative residual of the SC method (blue dots in Figure \ref{SCvsmDINMHConvergenceRate}) remains nearly constant until convergence at the 32nd iteration, except for a kink at the second iteration due to a huge residual at the first iteration. This behavior suggests a linear convergence rate for the SC method, which often requires a large number of iterations to converge.\\
In contrast, the relative residual of the mDINMH (orange dots in Figure \ref{SCvsmDINMHConvergenceRate}) shows a plateau until the 9th iteration, followed by an abrupt drop from the 10th iteration until convergence at the 13th iteration. This convergence pattern evidences superlinear convergence of the inexact Newton method enabled by the $Q$-order($1+p$) convergence condition (Equation \eqref{Qorder1+p}),\cite{holst_numerical_1995} which supports faster convergence of the mDINMH than the SC method as observed in Figure \ref{SCvsmDINMHConvergenceRate}. As a result, the mDINMH needs fewer iterations to achieve convergence, thereby significantly reducing the computational time as shown in Table \ref{SCvsmDINMHtime}.\\
\begin{figure}
    \centering
    \includegraphics[width=8cm]{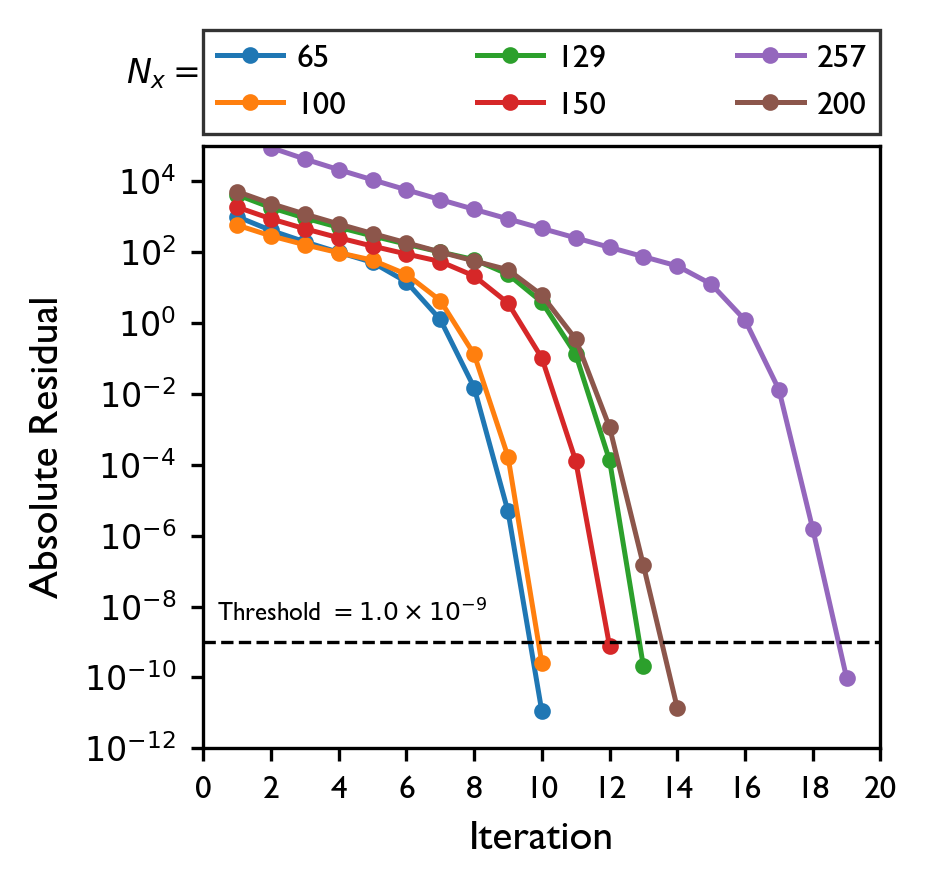}
    \caption{Absolute residual of the mDINMH method in solving the NLPBE for 4-nitroaniline calculated with various number of grid points. Calculations were performed with a 20 Å cubic box. The solute electron density is computed at B3LYP-D3/def2-TZVPPD level of theory. For $N_x$ = 257, NVIDIA A100-PCIE-40GB GPU is used for calculations due to the memory cost associated with the total grids.}
    \label{GridvsConv}
\end{figure}
\\
\textbf{Algebraic Multigrid.} We next discuss the applicability of the mDINMH to various numbers of grid points along $x$, $y$, and $z$-axis ($N_x$, See Figure \ref{FiniteDiff}) by examining their convergence according to iterative cycles, and the results are depicted in Figure \ref{GridvsConv}. Note that $N_x$ values in Figure \ref{GridvsConv} were carefully chosen to demonstrate the versatility of the AMG method across a broad range of grid points (See Section \ref{subsubsec:amg}).\\
Specifically, the GMG hierarchy in the original DINMH is contructed from the underlying grid structure, where the grid spacing is doubled at each coarser level and halved at each finer level when traversing the hierarchy (Figure \ref{GMGvsAMG}).\cite{holst_numerical_1995,coons_quantum_2018,womack_dl_mg_2018} Consequently, the original DINMH achieves optimal efficiency when $N_x = 2^n + 1$ ($n \in \mathbb{N}$), a condition satisfied by $N_x$ = 65, 129, and 257 in Figure \ref{GridvsConv}. In contrast, $N_x$ = 100, 150, and 200 in Figure \ref{GridvsConv} are even-numbered grid points, which the original DINMH is limited in its ability to handle.\\
Our mDINMH method, however, is capable of handling both cases by employing AMG as illustrated in Figure \ref{GridvsConv}. The absolute residual successfully reaches convergence within 20 iterations for all $N_x$ considered in Figure \ref{GridvsConv}, suggesting the stability of the mDINMH across a broad range of grid points enabled by AMG. \\
\begin{figure}
    \centering
    \includegraphics[width=8cm]{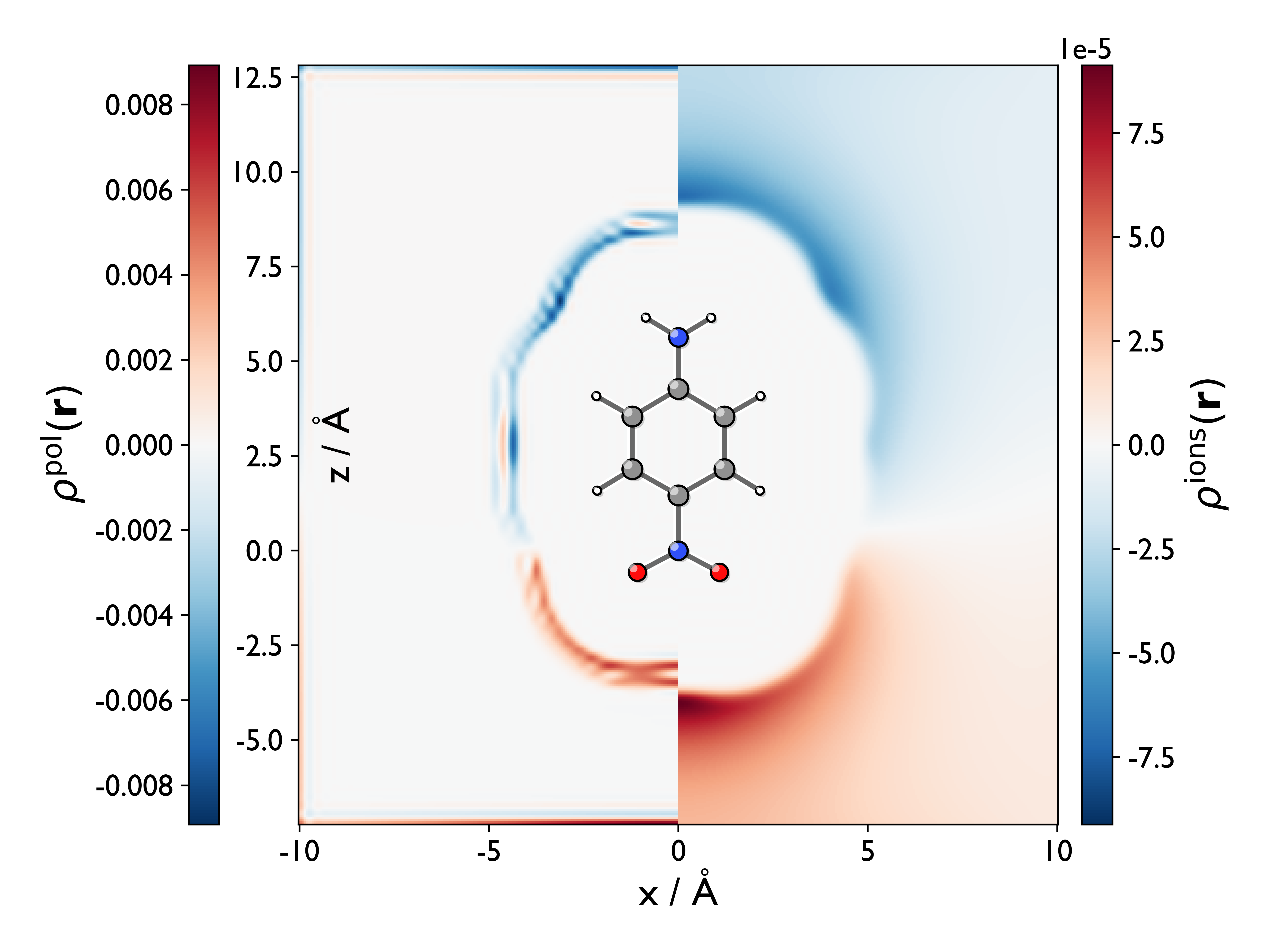}
    \caption{mDINMH-calculated $\rho^\text{pol}(\mathbf{r})$ and $\rho^\text{ions}(\mathbf{r})$ of 4-nitroaniline in 1 M of NaCl aqueous solution. Calculations were performed with a 20 Å cubic box discretized into 97$\times$97$\times$97 cubic grids. The solute electron density is computed at B3LYP-D3/def2-TZVPPD level of theory. Colorbars are in atomic unit.}
    \label{rho_pol_rho_ions}
\end{figure}
\\
\textbf{Polarization and Ion Charge Densities.} Based on the efficiency of the DF approach (Table \ref{AnalyticvsDensityFitting}) and the mDINMH method (Table \ref{SCvsmDINMHtime}), we consider their combination (DF+mDINMH) as our default method in solving the NLPBE. In this section, we first verify the qualitative accuracy of the DF+mDINMH method. Specifically, we examine $\rho^\text{pol}(\mathbf{r})$ (Equation \eqref{rho_pol}) and $\rho^\text{ions}(\mathbf{r})$ (Equation \eqref{MPB_ion_charge}), both of which are derived from the DF+mDINMH-calculated $\phi^\text{tot}(\mathbf{r})$. Both quantities were computed for 4-nitroaniline in 1 M NaCl aqueous solution, and the results are illustrated in Figure \ref{rho_pol_rho_ions}.\\
The calculated $\rho^\text{pol}(\mathbf{r})$ (Figure \ref{rho_pol_rho_ions}, left) exhibits a positively charged surface around the nitro group (–NO\textsubscript{2}). This positive $\rho^\text{pol}(\mathbf{r})$ is primarily attributed to the nitro-oxygen atoms, which adopt a partial negative charge in nitro moieties. These negatively charged oxygen atoms are expected to induce a positive charge in the environment to maintain local charge neutrality, resulting in a positive $\rho^\text{pol}(\mathbf{r})$ in their vicinity. Our speculation is supported by the restrained electrostatic potential (RESP) atomic charges (Table \ref{RESP}),\cite{bayly_well-behaved_1993} where the oxygen atoms carry a negative charge of –0.500. Similarly, the calculated $\rho^\text{ions}(\mathbf{r})$ (Figure \ref{rho_pol_rho_ions}, right) shows a positive value near the nitro moiety, suggesting the accumulation of cations caused by the negatively charged oxygen atoms.\\
In contrast, both $\rho^\text{pol}(\mathbf{r})$ (Figure \ref{rho_pol_rho_ions}, left) and $\rho^\text{ions}(\mathbf{r})$ (Figure \ref{rho_pol_rho_ions}, right) show negative values around the amino group (–NH\textsubscript{2}), which result from the positively charged amino-hydrogen atoms. The RESP charge of the these atoms is +0.383 (Table \ref{RESP}), suggesting that the hydrogen atoms develop a negative charge in the environment while increasing the local anion concentration around them.\\
Overall, the distributions of the mDINMH-calculated $\rho^\text{pol}(\mathbf{r})$ and $\rho^\text{ions}(\mathbf{r})$ are consistent with those reported by Head-Gordon and coworkers,\cite{stein_poissonboltzmann_2019} suggesting the reliability of our mDINMH method.\\
\begin{table}[t]
\centering
\caption{Restrained electrostatic potential atomic charges of 4-nitroaniline calculated at B3LYP-D3/def2-TZVPPD level of theory.}
\begin{minipage}{2cm}
    \centering
    \includegraphics[width=1.5cm]{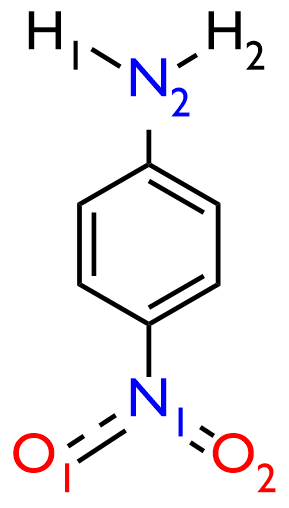}
\end{minipage}
\begin{minipage}{3cm}
    \centering
    \begin{tabular}{cc}
    \hline\hline
    Atom & RESP Charge \\
    \hline
    N\textsubscript{1} & +0.775\\
    O\textsubscript{1} & –0.500 \\
    O\textsubscript{2} & –0.500 \\
    N\textsubscript{2} & –0.863 \\
    H\textsubscript{1} & +0.383\\
    H\textsubscript{2} & +0.383\\
    \hline\hline
    \end{tabular}
\end{minipage}
\label{RESP}
\end{table}
\begin{table}
\caption{Electrostatic contribution to the free energy of solvation ($\Delta G_\text{es}^\text{solv}$, in kcal/mol) of 4-nitroaniline in 1 M NaCl aqueous solution calculated by four different computational schemes (self-consistent (SC)/mDINMH methods for solving the non-linear Poisson-Boltzmann equation and analytic/density fitting approaches for $\phi^\text{sol}_\text{elec}(\mathbf{r})$). Calculations were performed with a 20 Å cubic box discretized into 97$\times$97$\times$97 cubic grids. The solute electron density is computed at B3LYP-D3/def2-TZVPPD level of theory.}
\begin{tabular}{cccc}
\hline \hline
4-nitroaniline  & SC    & mDINMH & Q-Chem                 \\
\hline
Analytic        & –4.73 & –4.73  & \multirow{2}{*}{–4.74} \\
Density Fitting & –4.70 & –4.70  &                        \\
\hline \hline
\end{tabular}
\label{GsolvNitroaniline}
\end{table}
\begin{table}
\caption{Electrostatic contribution to the free energy of solvation ($\Delta G_\text{es}^\text{solv}$, in kcal/mol) of vitamin C in 1 M NaCl aqueous solution calculated by four different computational schemes (self-consistent (SC)/mDINMH methods for solving the non-linear Poisson-Boltzmann equation and analytic/density fitting approaches for $\phi^\text{sol}_\text{elec}(\mathbf{r})$). Calculations were performed with a 20 Å cubic box discretized into 97$\times$97$\times$97 cubic grids. The solute electron density is computed at B3LYP-D3/def2-TZVPPD level of theory.}
\begin{tabular}{cccc}
\hline \hline
Vitamin C       & SC    & mDINMH & Q-Chem                 \\
\hline
Analytic        & –5.26 & –5.26  & \multirow{2}{*}{–5.29} \\
Density Fitting & –5.26 & –5.26  &                        \\
\hline \hline
\end{tabular}
\label{GsolvVitaminC}
\end{table}
\\
\textbf{Solvation Free Energy.} While $\rho^\text{pol}(\mathbf{r})$ and $\rho^\text{ions}(\mathbf{r})$ can demonstrate the qualitative accuracy of the DF+mDINMH method, they are insufficient for validating its quantitative accuracy. Thus, we calculated the free energy of solvation (See Appendix \ref{FreeEnergy} for its derivation) to demonstrate the quantitative accuracy of the DF+mDINMH method, and the results are summarized in Table \ref{GsolvNitroaniline}. Specifically, the electrostatic contribution to the solvation free energy ($\Delta G^\text{solv}_\text{es}$) was computed for 4-nitroaniline in 1 M NaCl aqueous solution using four different computational schemes (analytic/DF for calculating $\phi^\text{sol}_\text{elec}(\mathbf{r})$ and SC/mDINMH for solving the NLPBE), and the calculated energies are compared with those from Q-Chem. See Section \ref{sec:ComputationalDetails} for computational details.\\
We first confirm the accuracy of the DF approach by comparing $\Delta G^\text{solv}_\text{es}$ computed with the analytic and the DF approaches (Table \ref{GsolvNitroaniline}). Both the SC and the mDINMH methods yield $\Delta G^\text{solv}_\text{es}$ of –4.73 for the analytic approach and –4.70 kcal/mol for the DF approach. Their difference is 0.03 kcal/mol only, which is negligible in practice. These results quantitatively confirm the accuracy of the DF approach.\\
Furthermore, Table \ref{GsolvNitroaniline} shows that the SC and the mDINMH methods produce the same $\Delta G^\text{solv}_\text{es}$ up to two digits irrespective to the computational approach for $\phi_\text{elec}^\text{sol}(\mathbf{r})$. Specifically, both methods produce –4.73 kcal/mol with the analytic approach and –4.70 kcal/mol with the DF approach. These agreements suggest the quantitative accuracy of the mDINMH method. The accuracy of the DF approach and the mDINMH method is further confirmed by $\Delta G^\text{solv}_\text{es}$ of vitamin C in 1 M NaCl aqueous solution (Table \ref{GsolvVitaminC}), where all four combinations results the same $\Delta G_\text{es}^\text{solv}$ of –5.26 kcal/mol.\\
We next evaluate the consistency of the DF+mDINMH method against the NLPBE solver in Q-Chem software (hereafter referred to as reference method).\cite{shao_advances_2015, stein_poissonboltzmann_2019} For 4-nitroaniline and vitamin C, the reference method produces $\Delta G_\text{es}^\text{solv}$ of –4.74 (Table \ref{GsolvNitroaniline}) and –5.29 kcal/mol (Table \ref{GsolvVitaminC}), while the DF+mDINMH results –4.70 (Table \ref{GsolvNitroaniline}) and –5.26 kcal/mol (Table \ref{GsolvVitaminC}). The deviations (0.04 and 0.03 kcal/mol for 4-nitroaniline and vitamin C, respectively) are practically negligible, suggesting that the DF+mDINMH is capable of producing reliable solvation free energies.\\
While the NLPBE has been developed primarily for modeling electrolyte solutions, it is also applicable to pure solvents by setting $c^\text{b} = 0$ M. This enables us to compare the NLPBE-calculated $\Delta G_\text{es}^\text{solv}$ with those from popular implicit solvent models, such as C-PCM,\cite{barone_quantum_1998} IEF-PCM,\cite{tomasi_ief_1999} COSMO,\cite{pye_implementation_1999} and SMD.\cite{marenich_universal_2009} For their rigorous comparison, we followed the standard PCM convention for the solvent-accessible surface, where atom-specific lengths are represented by scaled van der Waals radii $d_\text{A} = 1.2 \times d_\text{A,vdW}$.\\
Using these scaled van der Waals radii for solvent-accessible surfaces, we computed the solvation free energies of 4-nitroaniline and vitamin C in pure water using various implicit solvent models, and the results are summarized in Table \ref{GsolvPCM}. 
The C-PCM, IEF-PCM, and COSMO models produce $\Delta G_\text{es}^\text{solv}$ of approximately –15 and –23 kcal/mol for 4-nitroaniline and vitamin C, respectively (first to third row of Table \ref{GsolvPCM}). The SMD model (fourth row of Table \ref{GsolvPCM}) gives a comparable result of –14.98 kcal/mol for 4-nitroaniline, but predicts a less negative $\Delta G_\text{es}^\text{solv}$ of –16.18 kcal/mol for vitamin C, an outlier among the energies computed by standard solvation models considered in this study.\\
In contrast, the NLPBE-calculated $\Delta G_\text{es}^\text{solv}$ (fifth row of Table \ref{GsolvPCM}) is –32.00 and –53.15 kcal/mol for 4-nitroaniline and vitamin C, respectively. Compared to the standard solvent models, the NLPBE negatively overestimate $\Delta G_\text{es}^\text{solv}$ to a large extent. This overestimation is attributed to, unlike standard solvent models, smooth $\epsilon(\mathbf{r})$ at the solvent-accessible surface as defined in Equation \eqref{DielectricFunction}, which consequently allows a non-zero $\rho^\text{pol}(\mathbf{r})$ (See Equation \eqref{rho_pol} and Figure \ref{rho_pol_rho_ions}) inside the solute-accessible surface (See Figure \ref{SASIonExclusion}). Given the distance from the solute, this non-zero $\rho^\text{pol}(\mathbf{r})$ leads to strong electrostatic interactions with the solute charge density, thereby significantly contributing to the solvation free energy.\\
Therefore, we increased the scaling factor for atom-specific lengths from 1.20 to 1.45 ($d_\text{A} = 1.45 \times d_\text{A,vdW}$) to minimize such interactions and recalculated the solvation free energy. As summarized in the sixth row of Table \ref{GsolvPCM}, $\Delta G_\text{es}^\text{solv}$ is computed to be –14.77 and –21.51 kcal/mol for 4-nitroaniline and vitamin C, respectively. These values are  which are consistent with the values calculated by the C-PCM, IEF-PCM, COSMO, and SMD models, except for the SMD-calculated $\Delta G_\text{es}^\text{solv}$ for vitamin C as discussed above.\\
\begin{table}[t]
\caption{Electrostatic contribution to the free energy of solvation (in kcal/mol) of 4-nitroaniline and vitamin C in pure water calculated by the C-PCM, IEF-PCM, COSMO, and NLPBE (DF+mDINMH) implicit solvent models. The NLPBE calculations were performed with a 20 Å cubic box discretized into 97$\times$97$\times$97 cubic grids. The solute electron density is computed at B3LYP-D3/def2-TZVPPD level of theory.}
\begin{tabular}{ccc}
\hline \hline
     &  4-Nitroaniline & Vitamin C\\
\hline
 C-PCM    & –15.38 & –23.30 \\
 IEF-PCM & –15.28 & –23.06 \\
 COSMO & –15.21 & –23.08 \\
 SMD & –14.98 & –16.18 \\
 NLPBE\textsuperscript{a} & –32.00 & –53.15 \\
 NLPBE\textsuperscript{b} & –14.77 & –21.51 \\
 \hline \hline
\end{tabular}
\begin{flushleft}
\footnotesize
\textsuperscript{a} scaling factor = 1.20\\
\textsuperscript{b} scaling factor = 1.45
\end{flushleft}
\label{GsolvPCM}
\end{table}
We move on to ensuring the accuracy of the DF+mDINMH method by evaluating the effect of the electrolyte concentration to the solvation free energy of acetic acid \cite{stein_poissonboltzmann_2019} by calculating $\Delta \Delta G_\text{ions}$ defined as
\begin{equation}
    \Delta \Delta G_\text{ions} = \Delta G^\text{solv}_\text{es}(c^b) - \Delta G^\text{solv}_\text{es}(c^b = 0).
\end{equation}
We computed $\Delta \Delta G_\text{ions}$ with the DF+mDINMH and the reference method, and the results are depicted in Figure \ref{AceticAcid} (See Table S1 for raw data).
Similar to the work by Head-Gordon and coworkers (Figure \ref{AceticAcid}, Orange line), \cite{stein_poissonboltzmann_2019} both the DF+mDINMH and the reference method predict $\Delta\Delta G_\text{ions}$ as a non-linear function of electrolyte concentration (Figure \ref{AceticAcid}, Blue line), which consequently breaks the experimental linear trend (Figure \ref{AceticAcid}, Gray line).\cite{li_topological_2004} This deviation is well expected, as our theoretical framework also assumed a concentration-independent Stern layer thickness ($a$) when defining the ion-exclusion function ($\lambda(\mathbf{r})$, Equation \eqref{IonExclusion}).\cite{stein_poissonboltzmann_2019} Aside from this deviation, the DF+mDINMH-calculated $\Delta\Delta G_\text{ions}$ values (Figure \ref{AceticAcid}, Blue line) are in excellent agreement with those from the reference method (Figure \ref{AceticAcid}, Orange line),\cite{stein_poissonboltzmann_2019} confirming the numerical consistency of the DF+mDINMH combination with the reference method.
\begin{figure*}[t]
    \centering
    \includegraphics[width=18cm]{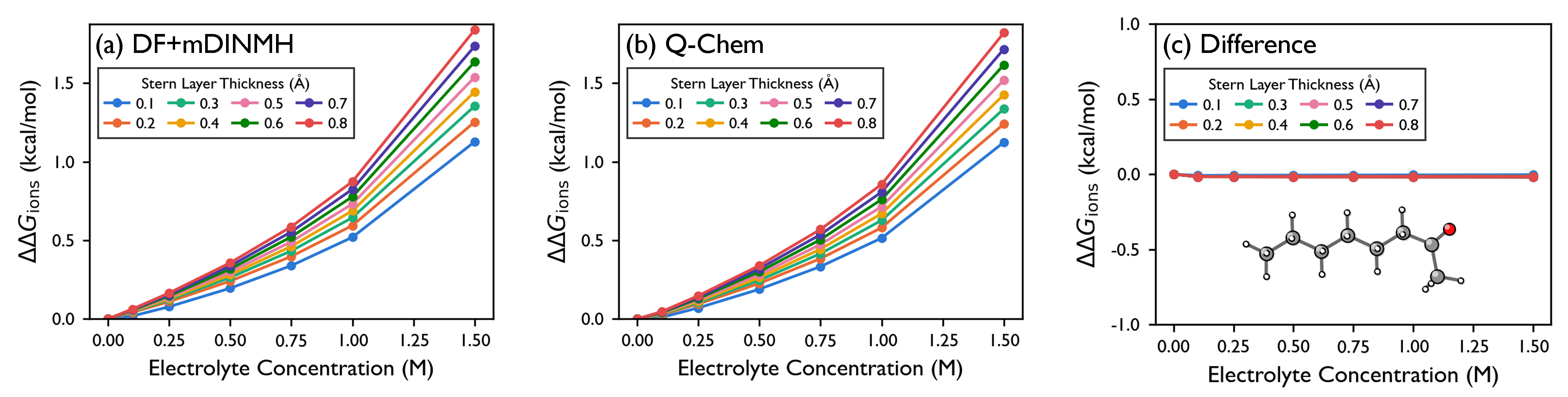}
    \caption{Calculated $\Delta\Delta G_\text{ions}$ (in kcal/mol) of 2-octanone with varying the Stern layer thickness and the concentration of NaCl aqueous solution with (a) the DF+mDINMH method, (b) the reference method available in Q-Chem software, and (c) their difference (See Table SX ). Calculations were performed with a 20 Å cubic box discretized into 97$\times$97$\times$97 cubic grids. The solute electron density is computed at $\omega$B97X-V/def2-TZVPP level of theory.\cite{mardirossian_b97x-v_2014}}
    \label{octanone}
\end{figure*}
To further validate the consistency of the DF+mDINMH method with the reference method, we computed $\Delta \Delta G_\text{ions}$ for 2-octanone with varying the Stern layer thickness (See Equation \eqref{IonExclusion}).\cite{stein_poissonboltzmann_2019} The results are depicted in Figure \ref{octanone} (See Table S2–S4 for raw data).\\
Consistent with the reference method (Figure \ref{octanone}b), our DF+mDINMH method predicts not only an increasing $\Delta \Delta G_\text{ions}$ as the electrolyte concentration increases but also the trend of $\Delta \Delta G_\text{ions}$ vs. electrolyte concentration curve with respect to the Stern layer thickness (Figure \ref{octanone}a). The difference between the DF+mDINMH- and the reference method-calculated $\Delta \Delta G_\text{ions}$ is almost negligible as depicted in Figure \ref{octanone}c.\\
In summary, we have demonstrated that the DF+mDINMH implemented in libNLPBE provides an efficient and accurate framework for predicting electrostatic potential profiles that molecules in electrolyte solutions experience. Therefore, we expect that libNLPBE serves as a powerful library for correcting electrostatic effects in density functional calculations.

\section{Example Usage}\label{sec:Examples}
We wrap up with providing simple hands-on examples to help users employ libNLPBE library for their own applications. Our libNLPBE library is primarily designed to be fully integrated into PySCF SCF calculations. Specifically, the library computes the solvation correction term that is incorporated into the Fock matrix during each SCF cycle. To facilitate this integration, we provide \verb|pbe_for_scf| function, which couples the solvent (\verb|cm|) and the SCF (\verb|mf|) objects by overriding the latter's \verb|kernel| function as shown in Listing \ref{lst:Listing1}.
\begin{code}
\begin{lstlisting}[language=Python]
from pyscf import gto
from libnlpbe.pbe import NLPBE, pbe_for_scf
mol = gto.M(atom='H 0 0 0; H 0 0 1')
mf = mol.RKS()
cm = NLPBE(mol, cb=1.0, length=10, ngrids=49)
cm.eps = 78.3553
solmf = pbe_for_scf(mf, cm)
solmf.kernel() # -1.0847550711994
\end{lstlisting}
\caption{Example of using libNLPBE for SCF calculations with PySCF.}
\label{lst:Listing1}
\end{code}
\\
For other quantum chemistry packages, libNLPBE supports single-shot NLPBE calculations to obtain the solvation free energy and the solvation correction term to the Hamiltonian as shown in Listing \ref{lst:Listing2}. This requires a \verb|molden| file containing information about molecular structure, orbital coefficients, and orbital occupancies, which are subsequently used to construct the solute density matrix (\verb|dm| in Listing \ref{lst:Listing2}) for calculating $\phi^\text{sol}_\text{elec}(\mathbf{r})$. Users can specify an auxiliary basis set when building \verb|NLPBE| object; otherwise libNLPBE uses def2-universal-JKFIT auxiliary basis set by default. The \verb|kernel| function of \verb|NLPBE| object returns the solvation free energy and the Hamiltonian correction term (\verb|epbe| and \verb|vmat| in Listing \ref{lst:Listing2}), which users can feed into their own code. Considering that most quantum chemistry codes support exporting SCF information into a \verb|molden| file, we expect libNLPBE to be applicable to a broad range quantum chemical platforms.
\begin{code}
\begin{lstlisting}[language=Python]
from pyscf.tools.molden import load
from pyscf.scf.hf import make_rdm1
f = load('test.molden')
mol, mo_coeff, mo_occ = f[0], f[2], f[3]
dm = make_rdm1(mo_coeff, mo_occ)
from libnlpbe.pbe import NLPBE
from pyscf.df.addons import make_auxmol
cm = NLPBE(mol, cb=1.0, length=10, ngrids=49)
cm.eps = 78.3553
cm.build(auxbasis='def2-universal-jkfit')
epbe, vmat = cm.kernel(dm=dm)
\end{lstlisting}
\caption{Example of using libNLPBE for single-shot NLPBE calculations.}
\label{lst:Listing2}
\end{code}

\section{Conclusion}
We developed libNLPBE, an open-source Python library for solving the non-linear Poisson-Boltzmann equation (NLPBE) for molecular systems. Our library features the density fitting (DF) approximation for calculating solute electrostatic potential, the mDINMH algorithm for efficiently solving the NLPBE, an algebraic multigrid (AMG) method to support various number of grid points, and GPU acceleration.\\
Using vitamin C and 4-nitroaniline as representative examples, we demonstrated that, compared to the analytic approach, the DF approach accelerates solute electrostatic potential calculations by $\sim$82 times on CPU and $\sim$302 times with GPU acceleration.\\
To account for the asymmetric Jacobian operator appearing in the Newton equation, we introduced a symmetric preconditioner in the mDINMH, which enables the application of efficient multigrid methods developed for symmetric operators. Unlike the original DINMH that requires symmetrization of $\nabla \epsilon(\mathbf{r}) \cdot \nabla$ through discretization, our symmetric preconditioner allows the mDINMH to use of the analytic gradient of $\epsilon(\mathbf{r})$, which is expected to reduce discretization errors. We showed that, compared to the self-consistent method for solving the NLPBE, the mDINMH algorithm achieves $\sim$10 and $\sim$74-fold speedups on CPU and GPU, respectively.\\
We examined the performance of the mDINMH across various numbers of grid points ($N_x$) to highlight AMG as a black-box method for solving the Newton equation. We showed that mDINMH calculations converged within 20 iterations for all $N_x$ considered in this study, confirming the applicability of the mDINMH to a broad range of grid counts.\\
Based on the performance of the DF approach and the mDINMH method, we propose a DF+mDINMH combination as an efficient computational scheme for solving the NLPBE. Its qualitative accuracy was examined by inspecting the distributions of the polarization charge density ($\rho^\text{pol}(\mathbf{r})$) and the ion charge density ($\rho^\text{ions}(\mathbf{r})$) around 4-nitroaniline. We showed that these are consistent with the calculated atomic charges, where the negatively charged nitro-oxygen atoms develop a positive $\rho^\text{pol}(\mathbf{r})$ and $\rho^\text{ions}(\mathbf{r})$ whereas the positively charged amino-hydrogen atoms do the opposite. We also compared $\Delta G_\text{es}^\text{solv}$ of 4-nitroaniline and vitamin C computed by the DF+mDINMH method at zero electrolyte concentration and popular solvent models such as C-PCM, IEF-PCM, COSMO, and SMD. The calculated results demonstrate that, after adjusting the scaling factor for atom-specific lengths, the DF+mDINMH method produces reasonable $\Delta G_\text{es}^\text{solv}$ compared to that from other solvent models.\\
We also examined the quantitative accuracy of the DF+mDINMH by calculating the electrostatic contribution to the solvation free energy ($\Delta G^\text{solv}_\text{es}$) of 4-nitroaniline and vitamin C. The calculated $\Delta G^\text{solv}_\text{es}$ under the DF approach differs by 0.03 and 0.00 kcal/mol for 4-nitrianiline and vitamine C, respectively, from those under the analytic approach. Furthermore, the mDINMH-calculated $\Delta G^\text{solv}_\text{es}$ does not show a difference from the SC method up to two decimal points, both of which confirm the accuracy of the DF+mDINMH method.\\
To further ensure its accuracy, we investigated the effect of electrolyte concentration on the solvation free energy using the DF+mDINMH approach. We found that the calculated free energy vs. electrolyte concentration curve is in great agreement with that produced by the reference method, confirming the accuracy of the DF+mDINMH method.\\
Our libNLPBE library can be fully integrated with PySCF SCF calculations. For other quantum chemistry code, libNLPBE offers single-shot calculations to obtain the solvation free energy and the solvation correction to the Hamiltonian, provided a \verb|molden| file that can be prepared by most quantum chemistry program. Therefore, we expect libNLPBE to be applicable to a broad range of quantum chemistry platforms.

\begin{acknowledgments}
J.-H.K. and W.Y. were supported by the American Chemical Society Petroleum Petroleum Research Fund (PRF\# 67056-ND6). We thank John M. Herbert (The Ohio State University) and Christopher J. Stein (TU Munich) for sharing their computational data and providing technical support for Q-Chem calculations.
\end{acknowledgments}

\section*{Data Availability Statement}
The source code of libNLPBE is available at https://github.com/Yang-Laboratory/libNLPBE.

\appendix
\section{Free Energy of Solvation}\label{FreeEnergy}
The non-linear Poisson-Boltzmann equation (Equation \eqref{NLPBE}) can be rewritten as the following expression
\begin{equation}
    \nabla \cdot \big ( \epsilon (\mathbf{r})\nabla \phi^\text{tot} (\mathbf{r})\big ) + 4\pi \big ( \rho^\text{sol}(\mathbf{r}) + \rho^\text{ions}[\phi^\text{tot}]\big ) = 0,
\end{equation}
which is a form of the Euler-Lagrange equation.\cite{sharp_calculating_1990} Thus, one can define the associated Lagrangian as below.
\begin{equation}
    L = 4\pi \rho^\text{sol}(\mathbf{r})\phi^\text{tot}(\mathbf{r}) - \cfrac{1}{2} \epsilon(\mathbf{r}) \big |\nabla \phi^\text{tot}\big |^2 - 4\pi \Delta \Pi [\phi^\text{tot}],
\end{equation}
where
\begin{equation}
    \cfrac{\delta \Delta \Pi [\phi^\text{tot}]}{\delta \phi^\text{tot}} = -\rho^\text{ions}[\phi^\text{tot}]
\end{equation}
The Euler-Lagrange equation is also satisfied for a functional $L' = C_1 L + C_2$, where $C_1$ and $C_2$ are a constant with respect to the variation of $\phi^\text{opt}$. We choose $C_1 = 1/4\pi$ for unit convention and $C_2 = 0$, such that
\begin{equation}
\begin{split}
    L' = &\rho^\text{sol}(\mathbf{r})\phi^\text{tot}(\mathbf{r}) - \cfrac{1}{8\pi} \epsilon (\mathbf{r}) |\nabla \phi^\text{tot}|^2 - \Delta \Pi[\phi^\text{tot}] 
\end{split}
\end{equation}
Thus, the electrostatic free energy is defined as
\begin{equation}
    G_\text{es} = \int \bigg [ \rho^\text{sol}(\mathbf{r}) \phi^\text{tot}(\mathbf{r}) - \cfrac{1}{8\pi} \epsilon(\mathbf{r}) |\nabla \phi^\text{tot}|^2 - \Delta \Pi[\phi^\text{tot}] \bigg ] d\mathbf{r}
\end{equation}
Now, we consider the electrostatic contribution to the solvation free energy.
\begin{equation}
\begin{split}
    G^\text{solv}_\text{es} &= G^\text{es}[\epsilon, c^b, \rho^\text{sol}] - G^\text{es}[1, 0, \rho^\text{sol}] - G^\text{es}[\epsilon, c^b, 0] \\
    &= \int \rho^\text{sol}(\mathbf{r}) \big ( \phi^\text{tot}(\mathbf{r}) - \phi^\text{sol}(\mathbf{r}) \big ) d\mathbf{r} \\
    &- \cfrac{1}{8\pi} \int \big ( \epsilon (\mathbf{r}) |\nabla \phi^\text{tot}|^2 - |\nabla \phi^\text{sol}|^2\big ) d\mathbf{r} - \int \Delta \Pi [\phi^\text{tot}] d\mathbf{r}
\end{split}
\end{equation}
Alternatively, the solvation free energy can be converted into the following form.
\begin{equation}
\begin{split}
    G^\text{solv}_\text{es} &= \cfrac{1}{2} \int \rho^\text{sol}(\mathbf{r}) \big ( \phi^\text{tot}(\mathbf{r}) - \phi^\text{sol}(\mathbf{r}) \big ) d\mathbf{r} \\
    &- \cfrac{1}{2} \int \rho^\text{ions} (\mathbf{r}) \phi^\text{tot}(\mathbf{r}) d\mathbf{r} - \int \Delta \Pi [\phi^\text{tot}] d\mathbf{r}
\end{split}
\end{equation}
For the size-modified ion charge density, $\Delta \Pi$ is given as below.
\begin{equation}
    \Delta \Pi[\phi^\text{tot}] = 2k_\text{B} T c_{1+2} \ln \bigg [ 1 + \cfrac{c^b}{c_{1+2}} \bigg ( \lambda(\mathbf{r}) \cosh ( \beta e\phi^\text{tot} ) - 1 \bigg )\bigg ]
\end{equation}

%\nocite{*}
\bibliography{references}% Produces the bibliography via BibTeX.

\end{document}